\documentclass[aps,pra,twocolumn,superscriptaddress,amssymb,nobibnotes,showkeys,showpacs]{revtex4-1}

\usepackage{graphicx}
\usepackage[ansinew]{inputenc}
\usepackage{array}
\usepackage{color}
\usepackage{amsmath}
\usepackage{amsxtra}
\usepackage{amstext}
\usepackage{amssymb}
\usepackage{float}
\usepackage{latexsym}
\usepackage{bm}
\usepackage[dvipsnames]{xcolor}
\usepackage{adjustbox}
\usepackage{multirow}
\usepackage{mathrsfs}
\DeclareMathOperator{\erf}{erf}

\newcommand {\Liao}[1]{ {\textcolor{black}{#1}}}

\begin{document}
	
	\title{Probing autoionizing states of molecular oxygen with XUV transient absorption: Electronic symmetry dependent line shapes and laser induced modification}	
	
	\author{Chen-Ting Liao}
	\affiliation{College of Optical Sciences and Department of Physics, University of Arizona, Tucson, AZ 85721, USA}
	
	\author{Xuan Li}
 \affiliation{Chemical Sciences Division, Lawrence Berkeley National Laboratory, Berkeley CA 94720, USA}
	
	\author{Daniel J. Haxton}
 \affiliation{Chemical Sciences Division, Lawrence Berkeley National Laboratory, Berkeley CA 94720, USA}
	
	\author{Thomas\ N. Rescigno}
 \affiliation{Chemical Sciences Division, Lawrence Berkeley National Laboratory, Berkeley CA 94720, USA}
	
	\author{Robert R. Lucchese}
	\affiliation{Chemistry Department, Texas A\&M University, College Station, TX 77842, USA}
	
	\author{C. William McCurdy}
	\email[]{cwmccurdy@lbl.gov}
 \affiliation{Chemical Sciences Division, Lawrence Berkeley National Laboratory, Berkeley CA 94720, USA}
  \affiliation{Department of Chemistry, University of California, Davis, CA 95616, USA}

	\author{Arvinder Sandhu}
	\email[]{sandhu@physics.arizona.edu}
	\affiliation{College of Optical Sciences and Department of Physics, University of Arizona, Tucson, AZ 85721, USA}	
	
	\date{\today}
	
\begin{abstract}

\Liao{We used extreme ultraviolet (XUV) transient absorption spectroscopy to study the autoionizing Rydberg states of oxygen in electronically and vibrationally resolved fashion}. XUV pulse initiates molecular polarization and near infrared (NIR) pulse perturbs its evolution. Transient absorption spectra show positive optical density (OD) change in the case of $ns\sigma_g$ {and $nd\pi_g$} autoionizing states of oxygen and negative OD change for $nd\sigma_g$ states. Multiconfiguration time-dependent Hartree-Fock (MCTDHF) calculations are used to simulate the transient absorption ~\Liao{and the resulting spectra and temporal evolution agree} with experimental observations. We model the effect of near-infrared (NIR) perturbation on molecular polarization and find that the laser induced phase shift model agrees with the experimental and MCTDHF results, while the laser induced attenuation model does not. We relate the electron state symmetry dependent sign of the OD change to the Fano parameters of the static absorption line shapes. 
		
\end{abstract}


\maketitle

\section{Introduction}
\label{sec:introduction}

{~\Liao{The interaction} of extreme ultraviolet (XUV) radiation with small molecules results in the formation of highly excited molecular states that evolve on ultrafast timescales and govern the dynamics of many physical and chemical phenomena observed in nature ~\cite{becker2012,ng1991}. In particular, single excitation of valence or inner valence electron to Rydberg molecular orbitals forms neutral states that lie above the ionization threshold (sometimes called ``superexcited'' states) ~\cite{platzman1962superexcited}. These states can lie energetically above several of the excited states of the molecular ion into which they can decay through autoionization ~\cite{Hatano1999}. Another feature of these autoionizing states is strong state-mixing and coupled electronic and nuclear motions, which can result in fast dissociation into excited neutral fragments ~\cite{nakamura1991basic}. The motivation for investigation of these states is quite wide ranging, from better understanding of~\Liao{the} solar radiation induced photochemistry of planetary atmospheres ~\cite{Wayne1991} to the~\Liao{ultraviolet} radiation damage in biological systems ~\cite{Bouda.2000.DNAdamage}. Furthermore, these are the states whose dynamics provide a mechanism for dissociative recombination of electrons with molecular ions, which has been the subject of decades of research~\cite{FlorescuMitchell2006,KokooulineGreene2011,Douguet_Orel_Greene2012,Kokoouline_Greene_Esry_2001,Guberman1991,Jungen_Pratt2010}. Due to their importance in complex processes, the direct observation of the electronic and nuclear dynamics of these states has been a topic of intense interest in molecular physics. }

Advances in ultrafast technology such as laser high-harmonic generation (HHG) have enabled femtosecond ($10^{-15}$ s) and attosecond ($10^{-18}$ s) light pulses in the energy range of 10-100's eV ~\cite{rundquist.1998.HHG, paul.2001.APTs}. These ultrashort and broadband XUV bursts provide a way to coherently prepare, probe, and control ultrafast dynamics of highly excited molecules ~\cite{gagnon.2007.MO,sandhu2008.O2}. Combined with time-delayed near infrared (NIR) or visible laser pulses, pump-probe spectroscopy schemes can be used to investigate dynamics in atoms and molecules on the natural timescale of electrons ~\cite{krausz2009}. 

To characterize excited state dynamics, researchers have developed sophisticated techniques involving the detection of charged photofragments,~\Liao{and the measurement of the photoabsorption signals}. In particular,  attosecond transient absorption spectroscopy (ATAS) has received considerable attention recently ~\cite{goulielmakis.2010.ATA.Kr, Wang.Chang.2010.ATA.Ar}, as it is relatively easy to implement and quite suitable for~\Liao{the} measurement and manipulation of bound and quasi-bound state dynamics. Recent ATAS experiments focus on properties and evolution of XUV initiated dipole polarization in atoms by using a delayed NIR pulse as a perturbation. In this scenario, many interesting ~\Liao{time-dependent phenomena} have been observed, including AC Stark shifts ~\cite{Chini.Chang.2012.ATA.ACstark}, light-induced states ~\cite{chen.2012.ATA.LIS}, quantum beats or quantum path interferences ~\cite{Holler.Keller.2011.ATA.WPI}, strong-field line shape control ~\cite{Pfeifer.2013.ATA.He.LorentzMeetsFano}, and~\Liao{resonant-pulse-propagation} induced XUV pulse reshaping effects ~\cite{LiaoPRL2015,LiaoPRA2016}. 

\begin{figure*}[ht!]
	\includegraphics[width=0.9\textwidth]{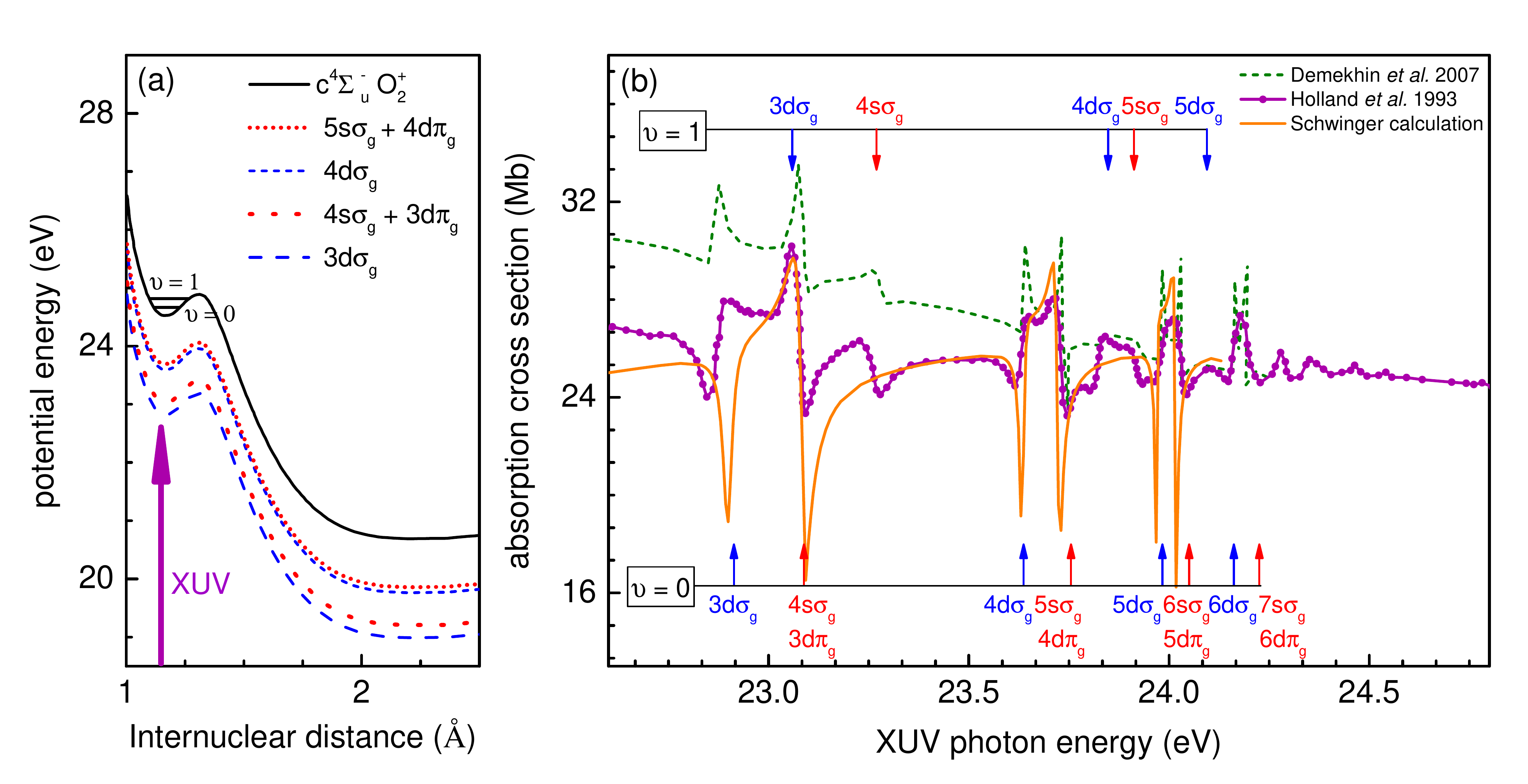}
	\caption{
		(Color online) (a) Schematic potential energy diagram of O$_2$. Black curve is the excited ionic c-state ($c^4\Sigma_u^{-}$). ~\Liao{Blue dashed and red dotted} curves are Rydberg series of neutral superexcited $nd$ and $ns$ states, respectively. Each state supports two vibrational levels. (b) (Purple line) Synchrotron measurement of photoabsorption cross section from ~\cite{Holland}. Features associated with various electronic states and their vibrational levels ($\nu$=0 at bottom and $\nu$=1 at top) are labeled.  (Green dash line) Theoretical photoabsorption cross section from ~\cite{Demekhin}. (Orange solid curve) Multichannel photoabsorption cross section that we obtained using Schwinger variational calculation. 
	}
	\label{FigPEcurves}
\end{figure*}

In contrast to bulk of previous studies, which have been conducted in atoms, our goal in this paper is to extend the ATAS for investigation of complex molecular systems. Molecular ATAS is a largely unexplored topic and very few studies have been conducted so far ~\cite{WarrickLeone.2016.ATA.N2,ReduzziSansone2016.ATA.N2}. ~\Liao{Here we present a joint experimental-theoretical study of the autoionizing Rydberg states of O$_2$.  An XUV attosecond pulse train was used to coherently prepare the molecular polarization, and a time-delayed NIR pulse to perturb its evolution. Superexcited states created by the XUV pulse have multiple} competing decay channels, including autoionization and dissociation into charged or neutral fragments. Autoionization process is one of the most fundamental process driven by electron correlation, which involves interference between ~\Liao{the} bound and continuum channels. This discrete-continuum interaction is ubiquitous in atoms, molecules and nano-materials ~\cite{NanoStructure.Fano.2010}, and it is formalized by well-known Fano formula describing their ~\Liao{spectral} line shapes ~\cite{Fano_1961}.


~\Liao{The paper is organized as follows. Sec. ~\ref{Autoionizingstates} below introduces the autoionizing Rydberg states of O$_2$ and compares the photoabsorption cross sections obtained by different methods, including our time-independent Schwinger variational calculations. In Sec. ~\ref{sec:TransientAbsorption_Experiment}, we describe our experimental setup and transient absorption line shapes obtained at various time delays. Section ~\ref{sec:TransientAbsorption_Theory} is focused on the theoretical approach, where we describe our time-independent Schwinger calculation method in subsection ~\ref{sec:Schwinger}, and the \textit{ab initio} MultiConfiguration Time-Dependent Hartree-Fock (MCTDHF) method in subsection ~\ref{sec:MCTDHF}. We then compare full experimental transient absorption spectrograms with MCTDHF calculations in subsection ~\ref{sec:comparison}. In Sec. ~\ref{sec:Model}, we present a simple model that connects Fano q parameter of static absorption profiles with the transient absorption line shapes and compares how laser induced attenuation (LIA) and laser induced phase (LIP) modifies the dipole polarization initiated by the XUV pulse. We summarize our work in Section ~\ref{sec:Conclusion}, followed by appendices that go into experimental details, MCTDHF approach, and few-level model of dipole polarization and NIR perturbation.}



\section{Autoionizing Rydberg states in O$_2$}
\label{Autoionizingstates}

\begin{table*}[htb!]
\centering

\caption{~\Liao{State assignment, effective quantum number n$^*$,  energy, linewidth and field-free lifetime of some relevant autoionizing states in O$_2$ from ~\cite{Holland, Demekhin}.  Fano $q$ parameters are obtained by fitting calculated photoabsorption cross section in Fig.~\ref{FigPEcurves}(b).\\ \\}}

\label{inset}
\begin{tabular}{cccccc}
    \hline
    \textbf{State} & \textbf{\quad n$^*$ \quad} & \textbf{Energy (eV)} & \textbf{Linewidth (meV)} & \textbf{Lifetime (fs)} & \textbf{Fano q} \\ \hline
    6s$\sigma_g$   & \multirow{2}{*}{$\sim$5} & 24.028               & 3.69                    & 178.37                   & -0.60          \\
    5d$\sigma_g$   &                          & 23.976               & 3.60                    & 182.83                   & 0.28           \\\hline
    5s$\sigma_g$   & \multirow{2}{*}{$\sim$4} & 23.733               & 7.31                    & 90.04                   & -0.59          \\
    4d$\sigma_g$   &                          & 23.632               & 7.10                    & 92.70                   & 0.22           \\ \hline
\end{tabular}
\end{table*}

In our experiment, autoionizing Rydberg states with electronic configurations that we can denote as 2s$\sigma_u^{-1}$( c$^4$$\Sigma_u^{-}$)nl$\sigma_g$, for example, are formed through the direct XUV excitation of an inner shell $2s\sigma_u$ electron to ~\Liao{the Rydberg} series converging to the excited ionic c-state ($c^4\Sigma_u^{-}$) of O$_2^+$. Fig.~\ref{FigPEcurves}(a) shows ~\Liao{the potential energy curve} of some of these states. For those states that are optically connected to the $^3\Sigma_g^-$ ground state of O$_2$ have the symmetries $^3\Sigma_u^-$ and $^3\Pi_u$. In our observations, those Rydberg series correspond to excitations from the 2s$\sigma_u$ orbital of O$_2$ to nd$\sigma_g$ ($\textsuperscript{3}\Sigma_u^-$), ns$\sigma_g$ ($\textsuperscript{3}\Sigma_u^-$),  and nd$\pi_g$ ($\textsuperscript{3}\Pi_u$). The XUV pulse in the experiment can also cause direct excitation to the $^3\Sigma_u^-$ and $^3\Pi_u$ continua by photoionization of the 3$\sigma_g$, 1$\pi_u$, and 1$\pi_g$ valence shells. Those excitations form the X$^2\Pi_g$, a$^4\Pi_u$, A$^2\Pi_u$ b$^4\Sigma_g^-$, B$^2\Sigma_g^-$ states of the ion,~\Liao{lying} below the c-state of O$_2^+$ ~\cite{Gilmore1965}, and also a second $^2\Pi_u$ state at 23.9 eV just below the ground vibrational level of the c-state at 24.564 eV ~\cite{Larsson1992}. 

Therefore, the autoionizing Rydberg states converging to the c-state are each embedded in several ionization continua, and can decay into any of them. These autoionizing Rydberg states can be grouped into pairs of dominant features ~\Liao{[nd$\sigma_g$, (n+1)s$\sigma_g$]} shown in Fig.~\ref{FigPEcurves}(a) as blue and red~\Liao{curves, respectively}. As we discuss later the ~\Liao{(n+1)s$\sigma_g$} series overlaps with~\Liao{the nd$\pi_g$} series, therefore, to be accurate, the pairs of features in Fig.~\ref{FigPEcurves}(a) should be listed as ~\Liao{[nd$\sigma_g$, (n+1)s$\sigma_g$+nd$\pi_g$]. The pairs relevant to our study are [3d$\sigma_g$, 4s$\sigma_g$+3d$\pi_g$], [4d$\sigma_g$, 5s$\sigma_g$+4d$\pi_g$], [5d$\sigma_g$, 6s$\sigma_g$+5d$\pi_g$], [6d$\sigma_g$, 7s$\sigma_g$+6d$\pi_g$]}, etc. 
Furthermore, the ionic c-state supports two vibrational levels ~\cite{Ehresmann2004.O2}, $\nu$=0 and $\nu$=1, as shown in Fig.~\ref{FigPEcurves}(a), and the nl$\sigma_g$ Rydberg states are also known to support at least two vibrational levels.

The pairs of Rydberg features~\Liao{[nd$\sigma_g$, (n+1)s$\sigma_g$+nd$\pi_g$]} can be identified in the static XUV photoabsorption spectra in Fig.~\ref{FigPEcurves}(b), where the purple curve is static absorption spectrum adapted from a synchrotron study by Holland \textit{et al.} ~\cite{Holland}. The~\Liao{autoionizing Rydberg series with vibrational state $\nu$=0 are labeled at the bottom of Fig.~\ref{FigPEcurves}(b)(blue and red labels for each pair), while series with vibrational state $\nu$=1 are labeled at the top}. The green curve in Fig.~\ref{FigPEcurves}(b) shows the theoretical cross section computed by Demekhin \textit{et al.} ~\cite{Demekhin} using a single center expansion method  that includes static and non-local exchange interactions without coupling between ionization channels leading to different ion states. These authors~\Liao{estimated} the $\nu$=1 contributions and broadened their theoretical cross sections by a Gaussian function of 20 meV full width at half maximum (FWHM).

Using the Schwinger variational method in calculations described in Sec. ~\ref{sec:Schwinger}, we computed the XUV photoionization cross section at the equilibrium internuclear distance of O$_2$ to approximate the vibrational ground state $\nu$=0 contribution.  Our results in Fig.~\ref{FigPEcurves}(b)(orange curve) reproduce the main features of synchrotron measurements very well. Our calculated cross section curve is for randomly oriented molecules and includes both perpendicular ($^3\Pi_u$) and parallel ($^3\Sigma_u^-$) contributions from various ionization channels corresponding to continua associated with different ionic states. From the calculated static absorption line shapes (orange curve), we extracted Fano $q$ parameters for a few representative states, which are listed in Table ~\ref{inset}. The autoionization lifetimes for these states \Liao{(based on Ref. ~\cite{Demekhin})} are also listed in Table ~\ref{inset}.
\begin{figure*}[htb!]
	\includegraphics[width=0.9\textwidth]{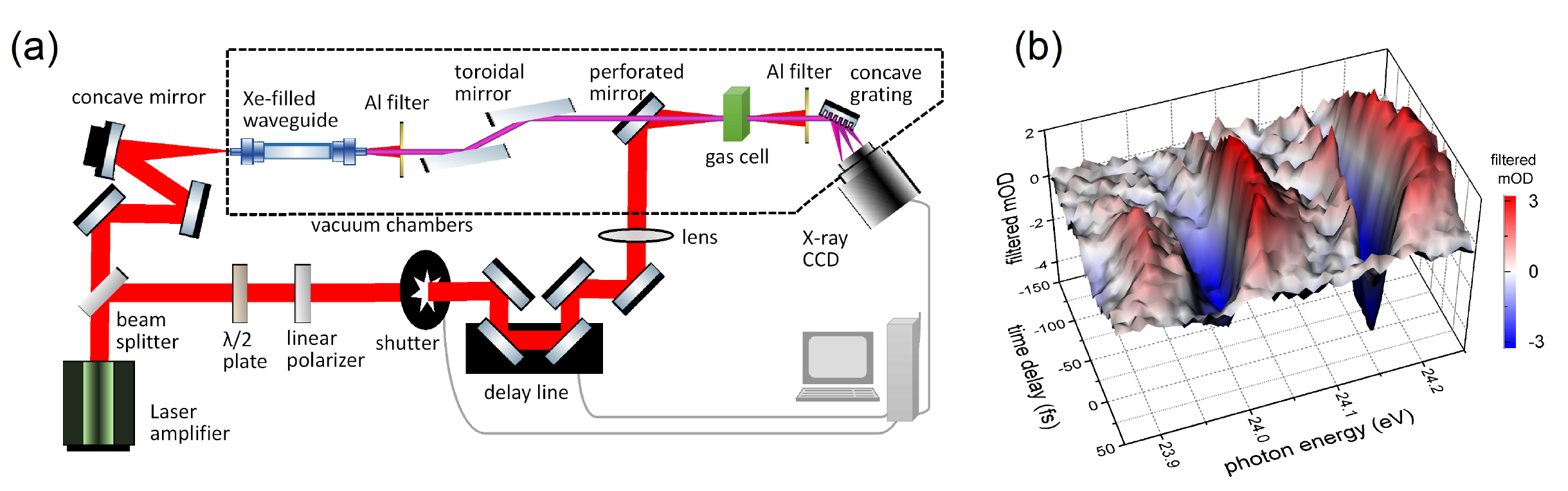}
	\caption{
		(Color online) (a) Experimental set up for XUV transient ~\Liao{absorption} studies in neutral superexcited states of O$_2$. (b) Experimentally measured transient absorption spectrogram in O$_2$.
	}
	\label{FigExpt} 
\end{figure*}

\section{Transient Absorption Experiment}
\label{sec:TransientAbsorption_Experiment}

To explore the dynamics of O$_{2}$ superexcited states, we conducted  experimental and theoretical ATAS studies. Our experimental pump-probe setup is shown in Fig.~\ref{FigExpt}(a).  Briefly, we employ 40 fs NIR pulses at 1 kHz repetition rate with pulse energy 2 mJ and central wavelength 780 nm. One portion of the NIR beam is focused into a xenon filled hollow-core waveguide to generate XUV attosecond pulse trains (APTs) with $\sim$440 attosecond bursts and $\sim$4 fs envelope. The APTs is dominated by harmonics 13, 15, and 17, out of which the 15th harmonic resonantly populates superexcited states. The second portion of NIR laser pulse goes through a delay-line and perturbs the XUV initiated molecular polarization with ~\Liao{estimated} peak intensity $\sim$1 TW/cm$^2$. \Liao{A grating spectrometer is used to measured the  XUV spectra transmitted through the O$_2$ gas sample. Using Beer-Lambert law we can determine optical density change (OD) due to NIR perturbation as a function of photon energy, $\hbar\omega$, and XUV-NIR time delay, $t_{d}$, as}
\begin{equation}
\label{eq:OD}
OD(\omega, t_d) = -\log[(I_{\tiny out}^{\scalebox{.5}{XUV+NIR}})/(I_{\tiny out}^{\scalebox{.5}{XUV}})],
\end{equation}
$I_{\tiny out}^{\scalebox{.5}{XUV+NIR}}(\omega, t_d)$ and $I_{\tiny out}^{\scalebox{.5}{XUV}}(\omega, t_d)$ are transmitted XUV spectra with and without the presence of NIR pulse, respectively. Further details of the experimental setup are given in Appendix \ref{sec:ExperimentalSetup}.

The spectrogram measured using ATAS is shown in Fig.~\ref{FigExpt}(b). To highlight the NIR induced absorbance change relative to continuum absorption, Fourier high pass filter is used to remove slow variation of underlying spectral profile, as also used in ~\cite{ott.2014.ATA.Recon, WarrickLeone.2016.ATA.N2}. Vertical axis of the spectrogram refers to milli- optical density change (mOD). Negative time delay means XUV arrives at the oxygen sample first, i.e. the NIR perturbation is imposed after the XUV initiates molecular polarization. There are many interesting aspects of this spectrogram. The striking feature being that we observe alternating blue and red bands corresponding to negative and positive OD change relative to the O$_2^+$ continua absorption spectrum, respectively. According to the  assignments of the autoionizing Rydberg states in Fig.~\ref{FigPEcurves}(b), we find that all nd$\sigma_g$ states show negative OD (less absorption compared to the continuum), while the features corresponding to the combination of ns$\sigma_g$ and nd$\pi_g$  show positive OD.

\begin{figure*}[htb!]
	\includegraphics[width=0.9\textwidth]{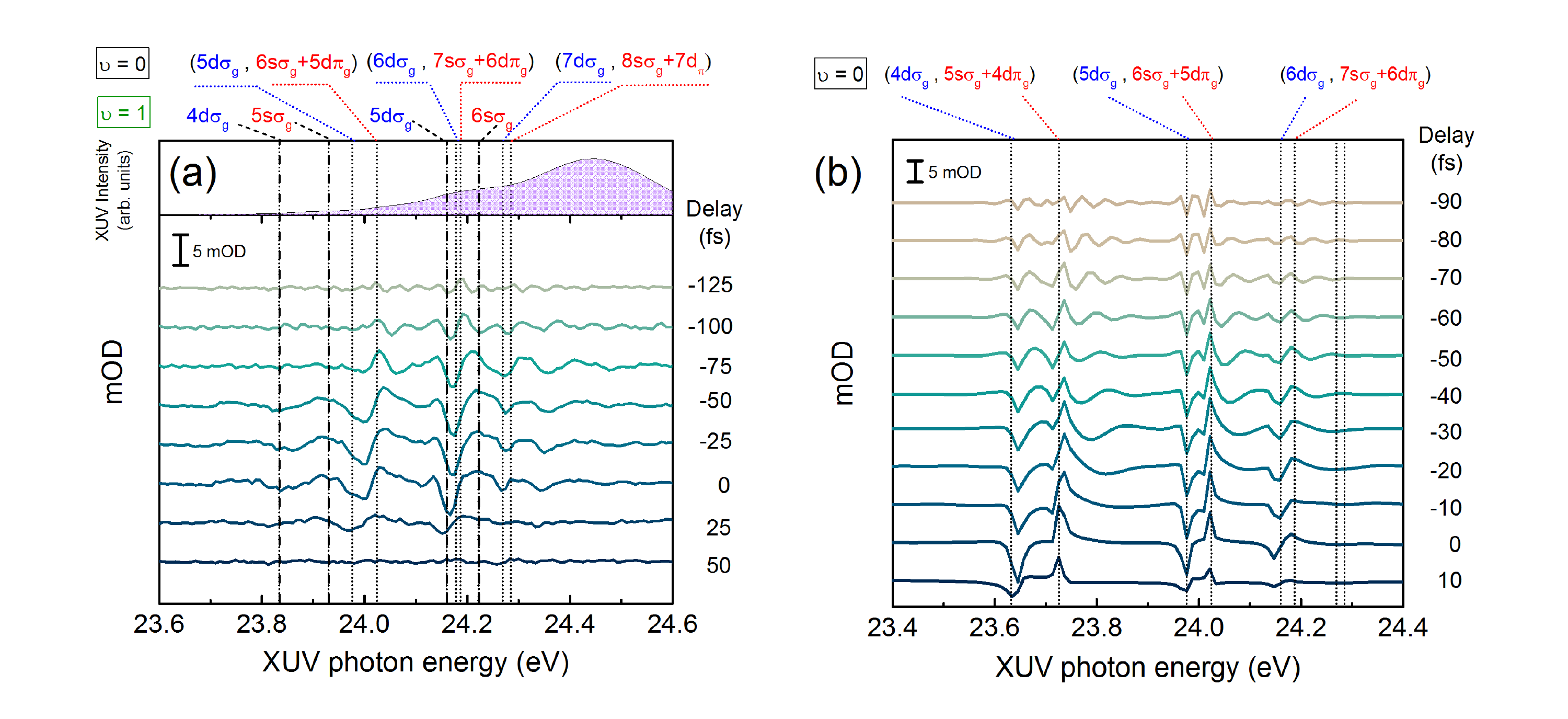}
	\caption{
		(Color online) (a) Experimental transient absorption spectra at certain representative time delays. \Liao{The scale bar represents 5 mOD change. Experimental XUV spectrum is also shown at the top.} Negative time delay implies XUV arrives at the target earlier than NIR pulse. All  nd$\sigma_g$ ~\Liao{(ns$\sigma_g$+nd$\pi_g$)} states show negative (positive) OD ~\Liao{at resonance}, corresponding to less (more) absorption compared to continua absorption spectrum. (b) The MCTDHF calculations of the transient absorption spectra at few time delays.
	}
	\label{FigATAspectra} 
\end{figure*}

We have plotted transient absorption spectra at some representative time delays in Fig.~\ref{FigATAspectra}(a). \Liao{The experimental input XUV spectrum is also plotted.} Relevant state assignments are labeled on the top of the figure. In addition to $\nu$=0 states, we also list $\nu$=1 states, following assignments in Ref. ~\cite{Demekhin}. 
In Fig.~\ref{FigATAspectra}(a), ~\Liao{for positive time delays, when NIR pulse arrives earlier,} there are no discernible features in the transient absorption spectrum, because the NIR pulse alone is not strong enough to significantly perturb the ground state of neutral O$_2$. At large negative delays where XUV arrives earlier than NIR, we observe finer oscillating structures corresponding to the well-known perturbed free induction decay ~\cite{wu.2016.ATA.review}. When the delay is close to zero, complicated line shapes can be observed. These line shapes are more complex than Fano profiles observed in ATAS of atomic gases. Considering different pairs of features in the series, i.e. [4d, 5s$\sigma_g$ + 4d$\pi_g$], [5d, 6s$\sigma_g$+ 5d$\pi_g$], etc, we find that all nd$\sigma_g$ features show a dip in the transient absorption line shape at the position of the resonance, while (n+1)s$\sigma_g$+nd$\pi_g$ show a peak. The difference in signs of OD for these features stems from the difference in Fano $q$ parameters of the static absorption profiles of the corresponding states which are shown in Fig.~\ref{FigPEcurves}(b) and discussed in more detail below. 

\section{Transient Absorption Theory}
\label{sec:TransientAbsorption_Theory}

\subsection{Time-independent calculations}
\label{sec:Schwinger}

~\Liao{Since the static XUV photoabsorption line shapes} play a central role in our interpretation of the transient absorption spectra, we calculated the XUV photoionization cross section using the Schwinger variational approach ~\cite{Stratmann1995,Stratmann1996}. 
~\Liao{Briefly,} the one-electron molecular orbitals in these calculations were expanded by using an augmented correlation-consistent polarized valence triple zeta aug-cc-pVTZ basis set ~\cite{Dunning1989,Kendall1992}. A valence complete active space self-consistent field calculation on the ground state of O$_2$ was used to obtain a set of orbitals that was then used in complete active space configuration interaction (CAS-CI) calculations on both the O$_2$ ground state and the O$_2^+$ states. The six channels that were included consisted of five channels which are open in the energy range of interest in this study, 23.6 eV to 24.4 eV,  $X\ ^2\Pi_g$, $a\ ^4\Pi_u$, $A\ ^2\Pi_u$, $b\ ^4\Sigma_g^-$,~\Liao{and} $B\ ^2\Sigma_g^-$.~\Liao{The} sixth channel, which was closed, is the $c ^{4}\Sigma _u^{-}$ channel that is responsible for the autoionization resonances studied here.  In all calculations, the ionization potentials were shifted slightly to agree with the experimental vertical ionization potentials~\Liao{in Ref. ~\cite{Baltzer1992}}.

\begin{figure*}[htb!]
	\includegraphics[width=0.8\textwidth]{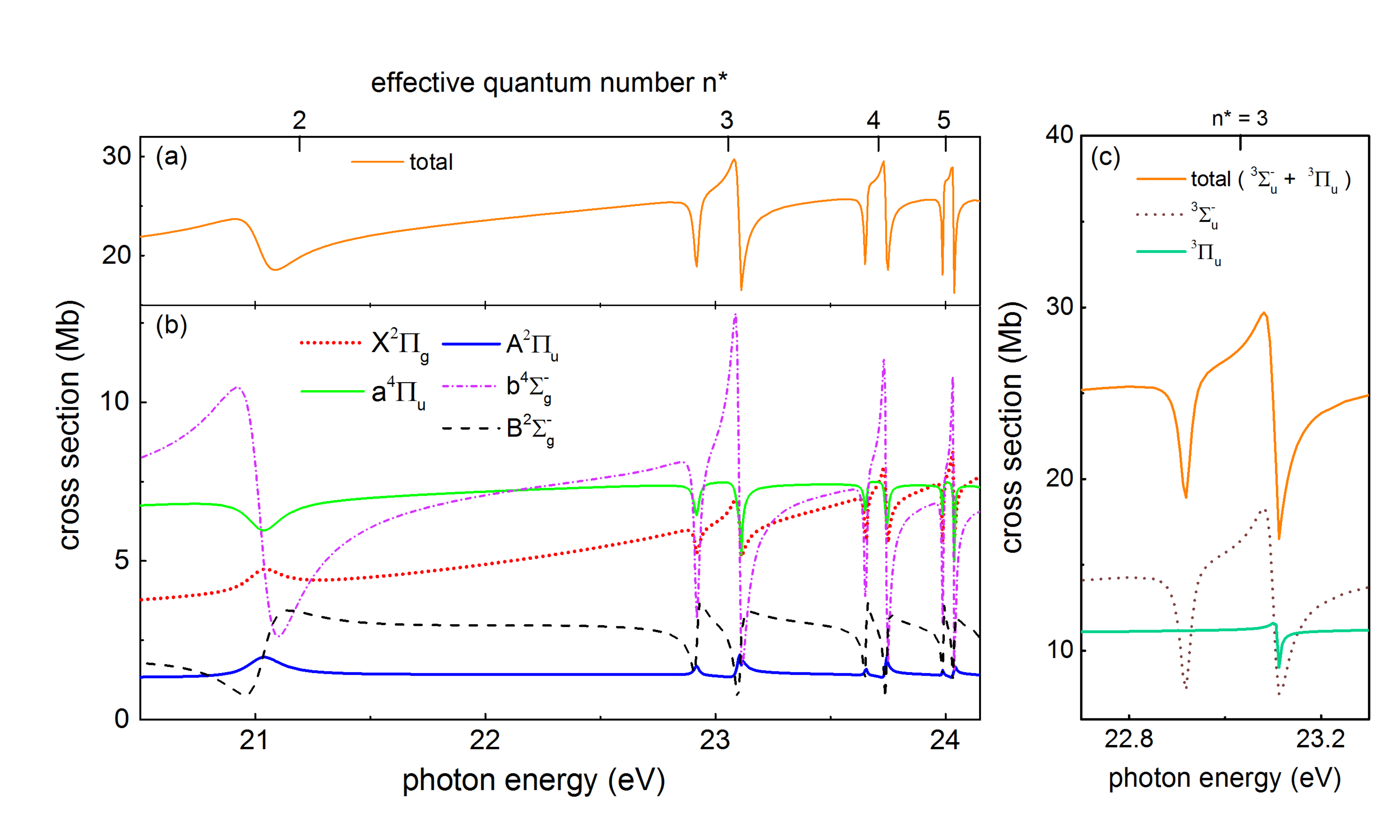}
	\caption{
		(Color online) (a) Total cross section as a function of the photon energy and effective quantum number n$^*$. (b) Various electronic channels contributing the total cross sections and their Fano profiles. (c) Parallel ($^3\Sigma_u^-$) and perpendicular ($^3\Pi_u$) polarization contributions to the total XUV photoionization cross section in the vicinity of the pair of features with n$^*$= 3 in Fig.~\ref{FigPEcurves} showing the presence of the ~\Liao{3d$\sigma_g$(parallel), 4s$\sigma_g$ (parallel), and 3d$\pi_g$ (perpendicular) states}.
	}
	\label{FigSchwinger} 
\end{figure*}

In Fig.~\ref{FigSchwinger}(a), we plot the total cross section as a \Liao{function of photon energy and as a function of the effective quantum number, defined as $n^{*}= \sqrt{R_{y}/(\rm{IP}-\hbar\omega)}$, where R$_y$ is the Rydberg constant and IP is the ionization potential of the closed $c ^{4}\Sigma _u^{-}$ channel which has the autoionizing resonances.} The partial cross sections computed for ionization leading to the five open channels are shown in Fig.~\ref{FigSchwinger} (b), which shows the autoionization resonances coming from the closed channel. We neglected two states that are observed in the photoelectron spectrum ~\cite{Baltzer1992},  \Liao{the dissociative $^{2}\Pi_u$ state at 23.9 eV that has a broad photoelectron spectra and the weak $^{2} \Sigma _u^{-}$ channel at 27.3 eV}. In addition, we have neglected a number of other experimentally unobserved states that would have very weak ionization cross sections in this energy region.

Fig.~\ref{FigSchwinger}(b) shows that Fano $q$ parameters depend on the final channel considered. However, a single resonance interacting with many continua can be rewritten as a resonance interacting with the linear combination of the channels. The orthogonal linear combinations of the channels do not interact with the resonance but contribute to a non-zero background to the cross section. Furthermore, by calculating the parallel ($^3\Sigma_u^-$) and perpendicular ($^3\Pi_u$) polarization contributions to the XUV photoionization cross section in Fig.~\ref{FigSchwinger}(c), we clearly see that each of the pairs of features in Fig.~\ref{FigPEcurves} corresponds to three states that were mentioned in the Sec. ~\ref{Autoionizingstates}.  ~\Liao{In each pair, the first feature} corresponds to an nd$\sigma_g$ Rydberg state while the second corresponds to contributions from the (n+1)s$\sigma_g$ and nd$\pi_g$ states. ~\Liao{The nd$\pi_g$ Rydberg series was proposed before by Wu \textit{et al.} ~\cite{Wu1987}, but it has been ignored in many of the subsequent experimental and theoretical studies.} Importantly, as seen from Fig.~\ref{FigSchwinger}(c), Fano $q$ parameters are similar for the two overlapping states (n+1)s$\sigma_g$ and nd$\pi_g$.

\subsection{Time-dependent calculations}
\label{sec:MCTDHF}

We performed \textit{ab initio} theoretical calculation of transient absorption signals in the same energy range as O$_2$ superexcited states using a recently developed implementation of the MCTDHF method. This method simultaneously describes stable valence states, core-hole states, and the photoionization continua, which are involved in these transient absorption spectra, and this approach has been previously explored and developed by several groups ~\cite{Cederbaum2007,Scrinzi_MCTDHF_2005,Kato_Kono2009,nest_lih2012,Miranda_2011, Madsen_2013,Sato_2013}. Briefly, our implementation solves the time-dependent Schr\"{o}dinger equation in full dimensionality, with all electrons active.  It rigorously treats the ionization continua for both single and multiple ionization using complex exterior scaling.   As more orbitals are included, the MCTDHF wave function formally converges to the exact many-electron solution, but here the limits of computational practicality were reached with the inclusion of full configuration interaction with nine ~\Liao{time-dependent} orbitals.  

 In these calculations,  an isolated XUV attosecond pulse is used to excite the polarization which is then perturbed by a more intense NIR pulse.  The weaker XUV probe pulse is modeled as an isolated 500 ~\Liao{attosecond} pulse with a sin$^2$ envelope centered at $27.21$ eV and with an intensity of $1.6\times10^{10}$ W/cm$^2$. The 40 fs NIR pulse is centered at 800 nm with an intensity of 1.5 $\times$ 10$^{12}$ W/cm$^2$.  The MCTDHF calculation produces the many-electron wave function $|\Psi(t)\rangle$ both during and after the pulses.

To describe the resulting spectrum, we start from a familiar expression for the transient absorption spectrum. If the time-dependent Hamiltonian is written as $\hat{H}=\hat{H_0}-\hat{d} \, \mathcal{E}(t)$ where $\hat{H_0}$ is the field-free Hamiltonian, $\hat{d}$ is the dipole operator, and $\mathcal{E}(t)$ is the electric field of the {applied XUV and NIR laser pulses}, the single-molecule absorption spectrum is proportional to the response function~\cite{TannorBook,Gaarde2011.ATA.Response,CDLin2013}, namely,
\begin{eqnarray}
\tilde{S}(\omega)=2\mbox{Im}[\tilde{d}(\omega) \, \tilde{\mathcal{E}}^*(\omega)]
\label{eq:ResponseFunc}
\end{eqnarray}

In this equation, $\tilde{d}(\omega)$ and $\tilde{\mathcal{E}}(\omega)$ are the Fourier transform of the time-dependent induced dipole, ${d}(t) =  \langle \Psi(t) | \hat{d} |\Psi(t) \rangle$, and the total applied electric field, $\mathcal{E}(t)$, respectively.   We use  response function in this study together with {\it ab initio} calculations of the electron dynamics to compute the transient absorption signals. Equation (\ref{eq:ResponseFunc}) is also the point of departure for our description of these spectra using the simple models described in Section~\ref{sec:Model}.  In both cases these spectra are used to compute the experimentally measured OD by employing the Beer-Lambert law as described in Appendix \ref{sec:MCTDHF_appen} where additional details of the MCTDHF calculations are given.

A number of previous calculations and experiments, for example, Refs. ~\cite{Demekhin,cubric1993.O2, liebel2000, liebel2002,hikosaka2003, doughty.Leone.2012.O2.VMI}, considered only parallel polarization between oxygen molecule internuclear axis and the XUV field, and thus invoked only two Rydberg series with ns$\sigma_g(^3\Sigma_u^-)$ (l=0, m=0) character and with $nd\sigma_g(\textsuperscript{3}\Pi_u)$ (l=2, m=0) character. In MCTDHF ~~\Liao{calculation}, as in the Schwinger variational calculation, we have assumed randomly oriented molecules in the presence of a linearly polarized XUV field in the calculation.  Here again therefore, in addition to Rydberg series corresponding to excitations form 2s$\sigma_u$ to ns$\sigma_g$ and nd$\sigma_g$ Rydberg orbitals, we ~\Liao{rediscovered} contributions from~\Liao{the} third Rydberg series corresponding to excitations to orbitals nd$\pi_g$ (l=2, m=1) character, converging to the same limit and forming O$_2$ states of overall $\textsuperscript{3}\Pi_u$ symmetry.  

Fig.~\ref{FigATAspectra}(b) shows MCTHDF calculation of the transient absorption spectra at few representative time delays. It agrees well with experimental spectra in Fig.~\ref{FigATAspectra}(a), {and all nd$\sigma_g$ states show a dip in the transient absorption line shapes, while (n+1)s$\sigma_g$+nd$\pi_g$ states show a peak}. Experimental line shapes exhibit more features than MCTDHF results due to the presence of transient absorption signals from additional vibrational level ($\nu$=1) for each state, and this possibility is not considered in the MCTDHF calculation. 

\begin{figure*}
	\includegraphics[width=0.9\textwidth,clip=true]{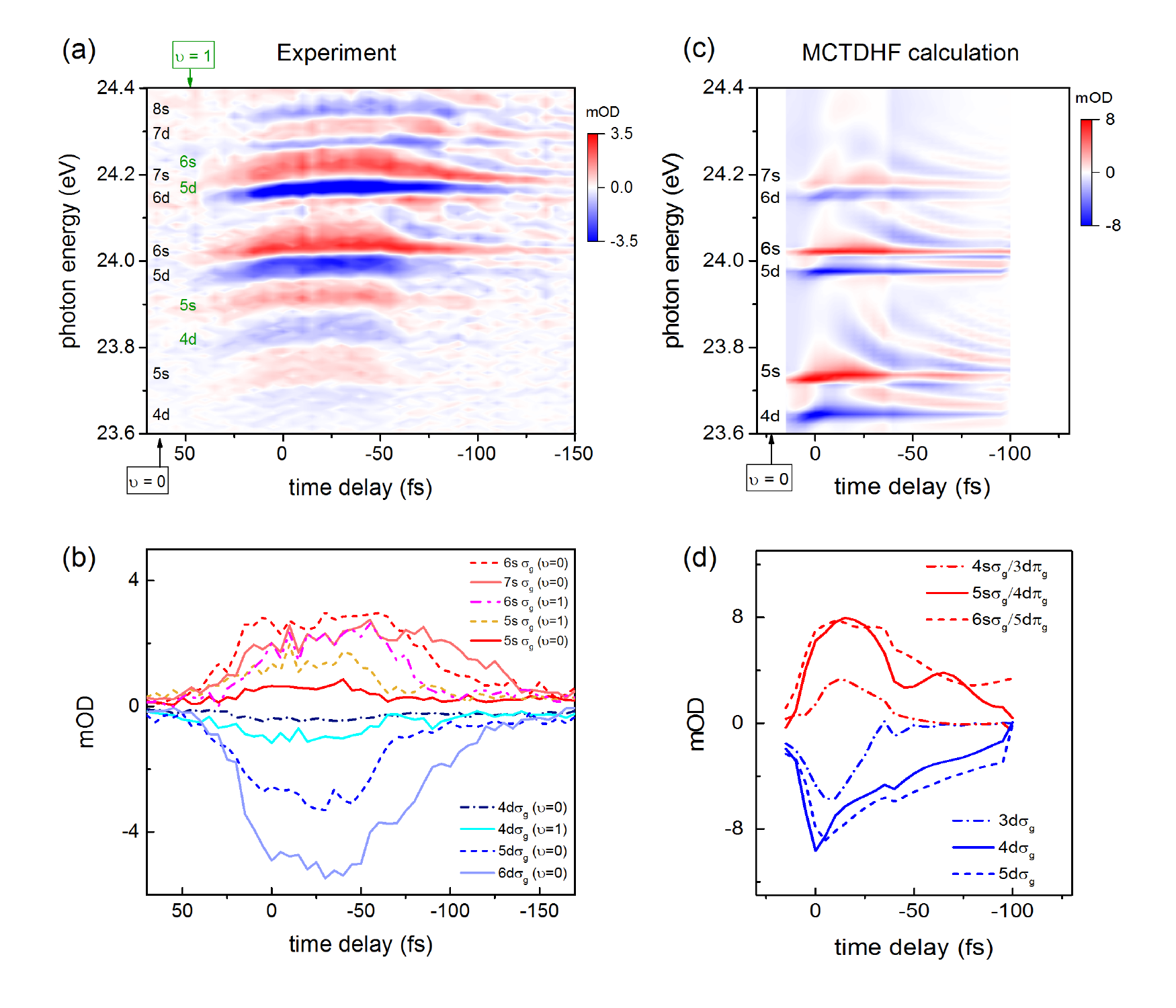}
	\caption{
		(Color online) (a) Measured transient absorption spectrogram labeled with various electronic states corresponding to $\nu$=0 (black) and $\nu$=1 (green) vibrational levels, and (b) its corresponding horizontal line-outs at some representative energy levels. (c) MCTDHF calculated transient absorption spectrogram, and (d) its corresponding horizontal line-outs.
	}
	\label{FigSpectrogram} 
\end{figure*}

\subsection{Comparison of Experimental and MCTDHF spectrograms}
\label{sec:comparison}

Next, we compare full experimental and calculated spectrograms as shown in Fig.~\ref{FigSpectrogram}(a) and (c), respectively. The MCTDHF calculation generally agrees with the experimental data very well, where they both show alternative positive (red) and negative (blue) absorbance structures at (n+1)s$\sigma_g$ {(including nd$\pi_g$)} and nd$\sigma_g$ states, respectively. Moreover, the upward curve of absorption structure indicates that there are AC Stark shifts of the quasibound states induced by moderate strong NIR laser field. The observed Stark shift $\sim$ 20 meV near zero delay corresponds well with the NIR laser peak intensity used. Also, hyperbolic fringes apparent at large negative time delays can be understood by the perturbed free induction decays, as also observed in ATAS studies in atomic gases. As mentioned earlier, experimental data contains contributions from $\nu$=1 vibrational levels, therefore the experimental spectrogram has more features than the theoretical counterpart.

By taking line-outs from the spectrograms at the energy location of various Rydberg states, we can obtain information about the evolution of these states. In Fig.~\ref{FigSpectrogram}(b), we take line-outs at the energies corresponding to the 4d($\nu$=0), 5s($\nu$=0), 4d($\nu$=1), 5s($\nu$=1), 5d($\nu$=0), 6s($\nu$=0), 6d($\nu$=0), 5d($\nu$=1), 7s($\nu$=0), 6s($\nu$=1), 7d($\nu$=0), 8s($\nu$=0) states, in order of increasing energy, at the position of maximum positive or negative OD change. Note that the line-outs labeled (n+1)s$\sigma_g$ include contribution from nd$\pi_g$. It is clear that regardless of the sign of the OD change the lifetime of polarization increases with the quantum number of the Rydberg series members. ~\Liao{It is known that the natural autoionization lifetimes scale with effective quantum number as $(n^{*})^3$ ~\cite{Lefebvre-Brion2004}, and for each n, the (n+1)s$\sigma_g$, nd$\pi_g$, and nd$\sigma_g$ states share the same $n^{*}$ (see Fig.~\ref{FigSchwinger}), and have similar decay timescales.} 

~\Liao{It should be noted that the decay timescales observed here are faster than the autoionization lifetimes due to NIR pulse induced broadening of resonances ~\cite{Li2015}. Fitting a convolution of Gaussian and exponential decay to the evolution of 5d, 6s ($\nu$=0) signals, we obtain a decay timescale $\sim$60 fs, corresponding to a net line width of $\sim$11 meV. Subtracting the field-free natural line width of 3.6 meV (Table ~\ref{inset}), we estimate the effective NIR induced broadening of these resonances to be $\sim$7 meV.  The decay timescales for other $\nu$=0 resonances are difficult to estimate as the transient absorption signals are either very weak or they overlap with vibrationally excited, $nu$=1 members of the Rydberg series. It is also hard to discern if decay timescales for $\nu$=1 states are different from the $\nu$=0 states.} This is significant as the dissociation lifetime of $\nu$=1 states are much shorter ($\sim$67 fs) than $\nu$=0 dissociation lifetime ($> ps$), and thus can be comparable to autoionization lifetime ~\cite{Padmanabhan2010.Lifetime} over the range of the effective quantum numbers considered here. One could argue in this case that as the molecule breaks up into excited atomic fragments, the decay of atomic polarization follows similar trend as the original molecular polarization. 

Fig.~\ref{FigSpectrogram}(d) shows MCTDHF calculation line-outs for certain ns($\nu$=0) and nd($\nu$=0) states, and their lifetimes trend qualitatively agrees with experimental observations and expectations that the larger effective quantum number state has a longer autoionization lifetime. However, our MCTDHF calculations are not able to accurately reproduce the absolute lifetimes of these states, particularly those with higher principal quantum number. {It should be noted that unlike some recent studies on excited states of $H_{2}$~\cite{Madsen2015_ATA_H2},  our experiment-theory comparison shows that for Rydberg autoionizing states in O$_2$ neither nuclear vibration nor molecular rotation has a significant effect on the delay-dependent line shapes obtained in transient absorption spectra.}



\section{Few-level Models for transient absorption spectra}
\label{sec:Model}

\begin{figure}[htb!]
	\includegraphics[width=0.25\textwidth]{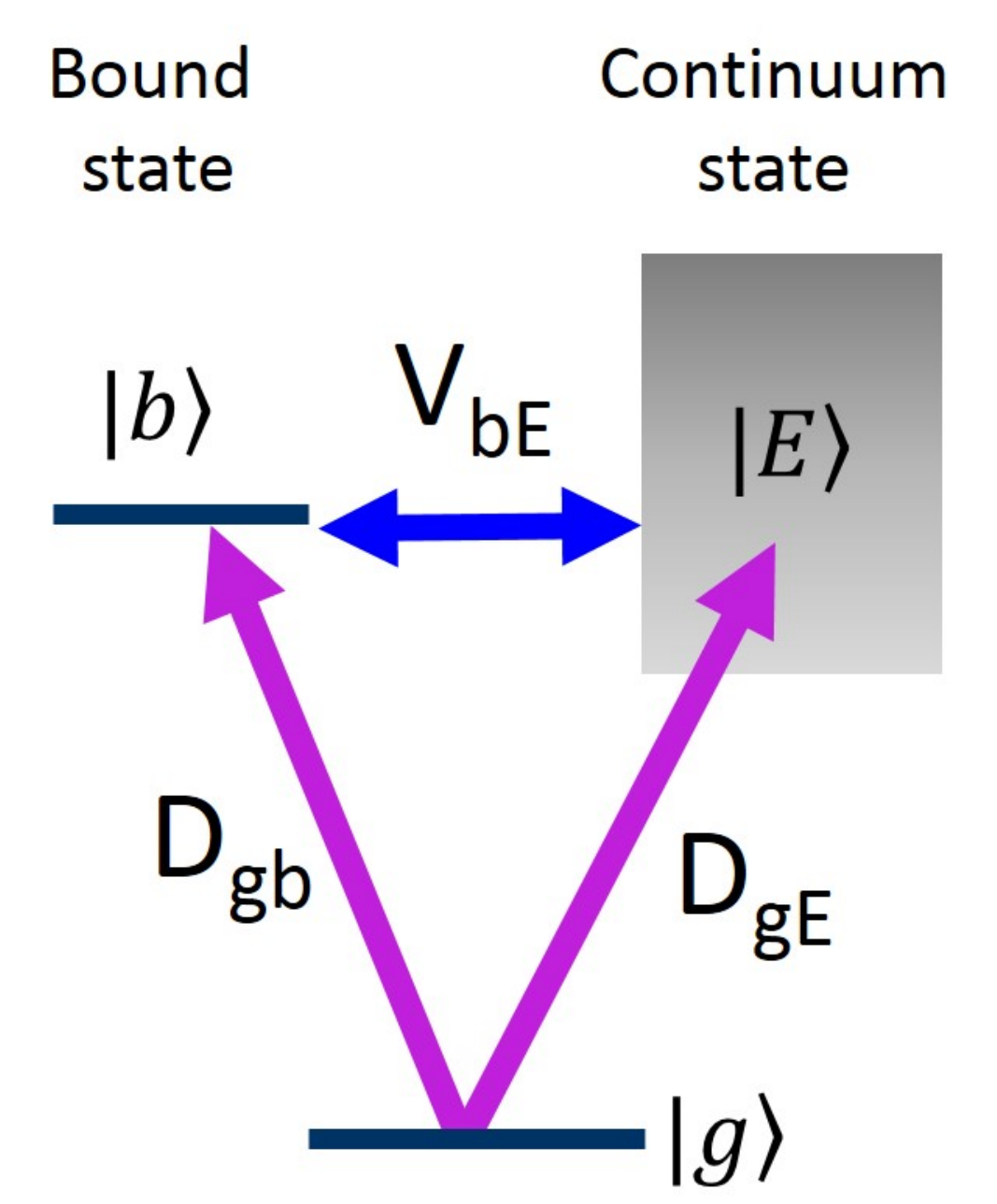}
	\caption{
		(Color online) ~\Liao{Three level model system showing a ground state $| g \rangle $, an excited bound state  $| b \rangle $, and a continuum state  $| E \rangle$, along with dipole couplings $D_{gb}$, $D_{gE}$, and interaction term $V_{bE}$.}
			}
	\label{FigThreeLevel} 
\end{figure} 

Transient absorption line shapes carry information on about the XUV induced dipole polarization and its modification by an NIR pulse.   For autoionizing states, we can use a few-level model to understand the origin of the correlation observed here between changes in the OD seen in transient absorption and Fano $q$ parameters of the line shapes in the corresponding static XUV absorption spectrum.  The central idea is to begin with a model for the polarization $d(t)$ induced by the XUV pulse that is capable of reproducing the Fano line shapes of the XUV spectrum, and then allow the NIR pulse ~\Liao{to} modify it by either laser-induced attenuation or laser-induced phase shift. ~\Liao{We then use modified $d(t)$ at time delay $t_d$, namely, $d_{t_d}(t)$, to calculate the change of the response function of Eq. (\ref{eq:ResponseFunc}), i.e. $\tilde{S}(\omega, t_d)-\tilde{S}(\omega)$,} and therefore the transient absorption spectrum.

In the model system, an XUV interaction directly couples the ground state, $| g \rangle $, to both an excited metastable quasibounded (autoionizing) state, $| b \rangle $, and the background continuum at nearby energies, $| E \rangle$.  A schematic of the energy levels ~\Liao{in our few-level model system} is shown in  Fig.~\ref{FigThreeLevel}. A similar treatment of autoionizing states using few-level model has been described in detail by Chu and Lin ~\cite{CDLin2013}, and we generally follow their approach. For the diagram in Fig.~\ref{FigThreeLevel}, the time-dependent wave function of the system is a superposition of these states
\begin{eqnarray}
     {\label{eq:total}}
\Psi (t)&=&e^{-iE_gt}C_g(t) | g \rangle \nonumber\\
&+&e^{-i(E_g+\omega_{\tiny \mbox{XUV}})t}\left[C_b(t) | b \rangle+\int dE C_E(t)| E \rangle \right].
\end{eqnarray}

The time-dependent coefficients $C_g(t)$, $C_b(t)$, and $C_E(t)$ can be computed by solving the corresponding time-dependent Schr\"{o}dinger equation in which the time-dependent Hamiltonian is 
\begin{equation}
{\label{Hamiltonian}}
\hat{H}(t)=\hat{H}_0-\hat{d}\left[\mathcal{E}_{\scalebox{.5}{XUV}}(t)+\mathcal{E}_{\scalebox{.5}{NIR}}(t; t_d)\right],
\end{equation}
where $\hat{H_0}$ is the unperturbed molecular Hamiltonian and $\hat d$ is the dipole operator.  The XUV field, $\mathcal{E}_{\scalebox{.5}{XUV}}(t)$ is centered at $t=0$, and the delayed NIR field is centered at $t=t_d$, so that
\begin{equation}
~\Liao{\mathcal{E}_{\scalebox{.5}{NIR}}(t;t_d) = \mathcal{E}^{0}_{\scalebox{.5}{NIR}} e^{{-(t-t_d)^2/\tau_{\scalebox{.5}{NIR}}^2}} e^{i \omega_L (t-t_d)} + c.c.,}
\end{equation}
where $\mathcal{E}^{0}_{\scalebox{.5}{NIR}}$ is NIR peak field amplitude, and $\hbar\omega_L$ is NIR photon energy.

We first consider the case with XUV pulse alone to verify that this simple treatment with two discrete levels and a background continuum can describe the Fano profiles of autoionizing states. The ultrashort XUV field creates a polarization, $d(t)$, at the beginning of the XUV pulse,
\begin{equation}
 \label{eq:dipole}
d(t)= \langle \Psi(t)| \hat d | \Psi(t) \rangle .
\end{equation}
We can solve for $|\Psi(t) \rangle $ and construct an analytic expression for $d(t)$ under the assumptions explained in Appendix \ref{sec:FewLevelModel}. We generalize the delta function XUV pulse used in ~\cite{CDLin2013} to have a Gaussian shape instead, {with envelope function $F(t)$ and electric field amplitude $F_{\scalebox{.5}{XUV}}$.}  
\begin{equation}
\begin{split}
&\mathcal{E}_{\tiny \mbox{XUV}}(t)=F(t) \exp{(i\omega_{\scalebox{.5}{XUV}} t)}+c.c. \\
&F(t) \equiv F_{\scalebox{.5}{XUV}}  \,\, e^{-(t/\tau_{\scalebox{.5} {XUV}})^2}/\sqrt{\pi \tau_{\scalebox{.5} {XUV}}^2}
\end{split}
\label{eq:GaussPulse}
\end{equation}
The analytical expression for $d(t)$ due to this XUV pulse is given in the appendix in Eq.(\ref{eq:DipoleExpression}). In our model calculations, the ~\Liao{the XUV pulse width $\tau_{\scalebox{.5} {XUV}}$ is 5 femtosecond}. In the limit that the XUV pulse is infinitely narrow ($\tau_{\scalebox{.5} {XUV}} \rightarrow 0$) this pulse becomes a delta function pulse as used in ~\cite{CDLin2013}.

\begin{figure*}
	\includegraphics[width=0.9\textwidth]{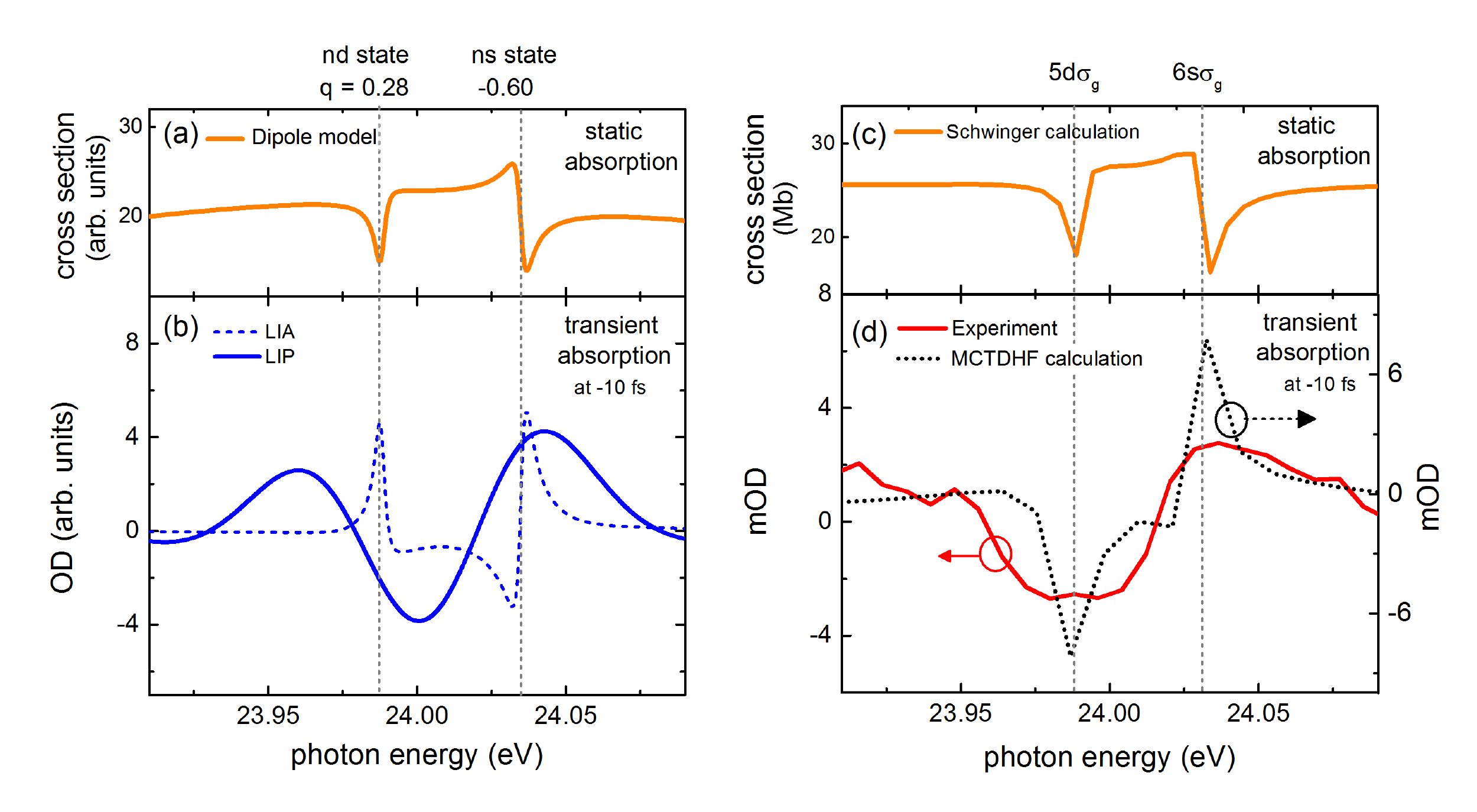}
	\caption{
		(Color online) Modeling of transient absorption line shapes resulting from the NIR perturbation. (a) Simulated static absorption profile that mimics the 5d and 6s state line shapes and the corresponding Fano $q$ parameters. (b) Transient absorption spectra obtained when NIR perturbation is modeled as laser induced phase (LIP) and laser induced attenuation (LIA). (c) Calculated static absorption spectrum for 5d and 6s states based on Schwinger variational approach ~\Liao{from Fig ~\ref{FigPEcurves}(b) for comparison}. (d) Experimentally measured transient absorption spectrum (solid red line) , and MCTDHF calculated transient absorption spectrum (dashed black line) ~\Liao{from Fig ~\ref{FigATAspectra}(b) for comparison}.
	}
	\label{FigModeling} 
\end{figure*}

In that limit this model produces precisely the Fano line shape of an absorption feature corresponding to the bound state, $| b \rangle$ embedded in the continuum as shown in Eq.(\ref{eq:Fano}).  Thus, our point of departure for the simple models for how the line shapes of the pure XUV spectrum are modified by transient absorption is the assumption that the NIR pulse arriving at time delay $t_d$ modifies the form of $d(t)$ initiated by the XUV pulse.  In these models, the decay lifetimes and Fano $q$ parameters are chosen so that the simulated frequency-domain static line shapes in Fig.~\ref{FigModeling}(a) for (5d,6s) states agree with our static photoabsorption cross section data in Fig.~\ref{FigPEcurves}(b).

To model the NIR perturbation of XUV induced polarizations in the atomic case, two approaches are widely used: the laser-induced attenuation (LIA) model and laser-induced phase (LIP) model. The key assumption of the LIA model is that the intense NIR pulse extinguishes the polarization initiated by ~\Liao{the} XUV pulse by truncating the oscillating electric dipole. Its physical meaning is that the quasibound state population is depleted by ~\Liao{the} NIR pulse ~\Liao{through transfer to} other states and continua. For sudden version of this approximation, the polarization can be expressed as
\begin{equation}
 \label{eq:LIA}
d_{t_d}(t)=\left\{
\begin{array}{cc}
d(t) &, t < t_d\\
0 &, t \ge t_d
\end{array}
\right. \quad \textrm{LIA model.}
\end{equation}
A slower truncation can also be applied by using a smoother function for transition between two regimes. The LIA approach is a central theme in a number of models for ATAS studies in many atomic gases~\cite{Bernhardt_Xenon_2014PRA, Pfeiffer_Leone_2013PRA, Li_Leone_JPhysB2015, CDLin2013}.

On the other hand, the LIP model assumes that NIR field leads to ~\Liao{an} energy modification, $\Delta E(t, t_d)$, of the excited state, and hence the polarization will gain additional phase such that 
\begin{equation}
 \label{eq:LIP}
d_{t_d}(t;t_d)=
d(t) e^{i\phi(t,t_d)} \quad \textrm{LIP model,}
\end{equation}
where ~\Liao{the} additional phase is $\phi (t, t_d) = \Delta E(t, t_d)/\hbar $, and it depends on the delay time $t_d$ of the NIR pulse. The LIP model has been used to explain results of ATAS in many atomic cases ~\cite{Mette.Ken.2013.ATA.He.LIP, wu.2016.ATA.review, Pfeifer.2013.ATA.He.LorentzMeetsFano,Chini.Chang.2012.ATA.ACstark}. 

~\Liao{Two methods can be used to calculate the energy and hence phase shift $\phi(t;t_d)$. Refs. ~\cite{Mette.Ken.2013.ATA.He.LIP,Pfeifer.2013.ATA.He.LorentzMeetsFano} calculate it based on the Stark energy shift, which is time-averaged and approximated as a pondermotive energy shift.} Alternately, second-order perturbation theory can be used to calculate the energy shift in the presence of coupling to nearby states ~\cite{Chini.Chang.2012.ATA.ACstark}.  Here we took the former approach ~\Liao{for the phase shift calculation,} and parameterized the pondermotive energy shift in atomic units as 
\begin{equation}
E_{\scalebox{.5}{pon}}(t;t_d) = [\mathcal{E}^{0}_{\scalebox{.5}{NIR}} e^{-(t-t_d)^2/\tau_{\scalebox{.5}{NIR}}^2}]^2 /(4 \omega_L^2).
\label{eq:LIPpon}
\end{equation}
Thus, the NIR modified dipole polarization becomes 
\begin{equation}
~\Liao{d_{t_d}(t;t_d) = d(t) e^{i E_{\scalebox{.5}{pon}}(t;t_d)\, t}.}
\label{eq:LIPdttd}
\end{equation}
Using either the LIP or LIA ~\Liao{model, we Fourier transformed the modified polarization and calculated the response function using Eq.(\ref{eq:ResponseFunc}), thereby obtaining the OD as a function of energy and delay using Eq.(\ref{eq:opticaldensity}).  Further details of our model calculations are given in Appendix ~\ref{sec:FewLevelModel}}.

In the LIA model, we use smooth truncation of the XUV initiated polarization, which results in the absorption spectrum shown as dashed blue line of Fig.~\ref{FigModeling}(b) (at delay -10 fs).  In this model, the absorption at resonance energy increases or decreases depending on whether Fano $q$ parameter of the static line shape is either less than or greater than unity, similar to the behavior for a sudden truncation case exhibited explicitly in Eq. (\ref{eq:suddenLIA}). The Fano $q$ parameters of the pairs of autoionizing features in ~\Liao{static} absorption spectrum (Fig.~\ref{FigPEcurves}(b)) both have $q <1$, but with differing signs, and thus the LIA model, even with gradual attenuation, fails to reproduce the directions of absorption changes observed in the transient absorption spectrum.

Using Fano $q$ parameters from the static absorption profile in Fig.~\ref{FigModeling}(a) and our experimental NIR pulse parameters, we also applied the LIP model with pondermotive energy shift as a NIR perturbation effect, and results are shown in solid blue line in Fig.~\ref{FigModeling}(b).
The application of a laser induced phase evidently produces an effect in which the absorption increases or decreases depending on the \textit{sign} of $q$ when $q < 1$. For comparison of our model with full theory and the experiment, we show static absorption line shapes obtained using Schwinger calculation in Fig.~\ref{FigModeling}(c). Figure~\ref{FigModeling}(d) shows the transient absorption line shapes obtained experimentally (solid red line) and from MCTDHF calculation (solid black line) at delay of -10 fs. Both our experimental and MCTDHF calculated line shapes agree well with the LIP model. It is important to note that the detailed measurements of complex line shapes associated with molecular polarization of pairs of nd and ns states provide the required detail to distinguish between the validity of different models. To our knowledge, ~\Liao{this} is the first study where such comparison has been made. Based on our results, it seems that in this case the LIP assumptions better reflect the physics of ATAS experiment than those of the LIA model, even with smooth attenuation functions. 

\section{Conclusion}
\label{sec:Conclusion}

In summary, we used the ATAS to investigate XUV initiated oxygen molecular polarization of superexcited states, perturbed by a NIR pulse, and the alternate negative and positive absorption spectra for nd$\sigma_g$  and (n+1)s$\sigma_g$ autoionizing states are observed. The numerical results obtained using {\it ab initio} MCTDHF calculations agree with the experimental findings. ~\Liao{In addition, from} our MCTDHF and Schwinger variational calculations, we identify and include the contribution of a weaker nd$\pi_g$ state that overlaps with (n+1)s$\sigma_g$ state. From the transient absorption spectrograms, we observe that decay lifetime of the dipole polarization for nd$\sigma_g$ and (n+1)s$\sigma_g$ states is similar and it increases with the effective quantum number ~\Liao{n$^*$. However, the decay timescale is faster than natural autoionization timescale due to NIR pulse induced broadening of resonances.} The decay lifetime is also found to be insensitive to the vibrational state of the molecule, within the sensitivity of our measurements. ~\Liao{To better interpret our findings}, two models of NIR perturbation of the XUV initiated molecular polarization are tested against experimental and MCTDHF calculated transient absorption line profiles, and we find that laser induced phase shift model explains our results, while laser induced attenuation does not. On these grounds, we conclude that the negative/positive transient absorption signals for nd/ns states can be explained in terms of two very different manifestation of electronic interference in molecular excitation (opposite signs of initial Fano $q$ parameters) influenced by the same amount of NIR induced Stark shift in transient absorption experiments. 
We envision that additional ATAS investigations of {low quantum number Rydberg states that do not follow core-ion approximation ~\cite{hikosaka2003}, with few-cycle NIR pulses, will enable us to study the non-adiabatic effects associated with fast autoionization and dissociation in O$_2$. ~\Liao{The relationship between the static properties and transient absorption line shapes explored here leads us to propose that finer features of ATAS spectra in molecules could be used to characterize the undetermined electronic properties of dynamically evolving systems} and test the theoretical models of the strong-field modification of the correlated electron dynamics}, including recently proposed interference stabilization of autoionizing states ~\cite{EcksteinVrakking2016.N2}.

\begin{acknowledgments}
	Work at the University of Arizona and the University of California Davis
	was supported by the U. S. Army Research Laboratory and the U. S. Army Research Office under grant number W911NF-14-1-0383.	Work performed at Lawrence Berkeley National Laboratory was supported by the US Department of Energy Office of Basic Energy Sciences, Division of Chemical Sciences Contract  DE-AC02-05CH11231. C.-T.L. acknowledges support from Arizona TRIF Photonics Fellowship. C.-T.L and X.L. contributed equally to this work in the form of experimental and theoretical effort, respectively. 
\end{acknowledgments}



\appendix
\section{Experimental Setup}
\label{sec:ExperimentalSetup}

A Ti:Sapphire laser amplifier is used to produce 40 fs NIR pulses at 1 kHz repetition rate with pulse energy 2 mJ, central wavelength 780 nm, with no active control of carrier envelope phase. After exiting the amplifier, the NIR pulse is divided into two paths. The NIR pulse ~\Liao{on the} first path is focused into a xenon gas filled hollow-core capillary waveguide to generate XUV APT with $\sim$440 attosecond bursts and $\sim$4 fs envelope via HHG process. The APT is dominated by harmonics 13, 15, and 17. The harmonic XUV beam is passed through ~\Liao{an} aluminum filter to ~\Liao{remove} residual NIR, and then a toroidal mirror is used to focus ~\Liao{the XUV} into a gas cell, which constitutes our interaction region. The 15th harmonic in the XUV beam is resonant with neutral superexcited states of oxygen and initiates the molecular polarization. ~\Liao{The delayed NIR pulse on the second path passes through a focusing lens, and} it is recombined collinearly with XUV beam using a mirror with a hole.  Both XUV and NIR pulses impinge on a 1 cm length oxygen gas cell with a backing pressure of 4 torr, with aluminum ~\Liao{foils} providing gas to vacuum partition. The NIR pulse, with focused peak intensity at $\sim$1 TW/cm$^2$, drills through covering ~\Liao{foils}, allowing both XUV and NIR beams to propagate forward collinearly towards an XUV spectrometer.

A home-made ~\Liao{XUV} spectrometer is used, which includes a concave grating (1200 lines/mm, 1 m radius of curvature) and a back-illuminated thermoelectric-cooled X-ray CCD camera. Another 200 nm thick aluminum filter is equipped in front of the camera to block NIR. We use a shutter in the NIR delay line to obtain background (NIR free) XUV only spectra $I_{\tiny out}^{\tiny XUV}$ at each camera exposure. The spectrometer detects transmitted XUV spectra with a resolution of $\sim$10 meV at 24 eV. The spectrometer does not resolve the narrow NIR-free oxygen absorption lines, therefore the transmitted XUV spectrum $I_{\tiny out}^{\tiny XUV}$ in the absence of NIR field is essentially the same as the input XUV spectrum $I_{\tiny in}^{\tiny XUV}$. We use it as a reference in OD measurements. The experimental OD is obtained from near-simultaneously measured transmitted XUV spectrum with NIR present, and without NIR, with 0.1 s exposure time per camera exposure.  \Liao{The absolute values of experimental OD are shifted slightly lower due to the presence of residual camera background in the raw data, and this effect can be significant at photon energies where XUV intensity is very low. We averaged 200 camera frames at each delay step and the statistical errors bars on our data range from $\pm$1-2 mOD.}

\section{MCTDHF method}
\label{sec:MCTDHF_appen}

To calculate the transient absorption spectra, we applied the MCTDHF method, which simultaneously describes stable valence states, core-hole states, and the photoionization continuua. MCTDHF implementation solves the time-dependent Schr\"odingier equation in full dimensionality, and because it is based on a combination of the discrete variable representation (DVR) and exterior complex scaling (ECS) of the electronic coordinates, it rigorously treats the ionization continua for both single and multiple ionization. As more orbitals are included, the MCTDHF wave function formally converges to the exact many-electron solution. The MCTDHF electronic wave function is described by an expansion in terms of time-dependent Slater determinants, $| \Psi(t) \rangle =\sum_a A_a(t) | \vec{n}_a(t) \rangle,$ in which each determinant is the anti-symmetrized product of N spin orbitals, $ | \vec{n}_\alpha(t) \rangle =\mathscr{A}\left( | \phi_{a1}(t) \rangle \cdots | \phi_{aN}(t) \rangle \right)$. These spin-restricted orbitals $ | \phi_{a}(t) \rangle $ are in turn expanded in a  set of  time dependent discrete DVR basis functions, and full configuration interaction is employed within the electronic space. We reach the MCTDHF working equations by applying the Dirac-Frenkel variational principle to the time-dependent Schr\"odinger equation for the trial function. Details of the resulting working equations and their solution can be found in Ref.~\cite{HLM2011}.

The results presented here were calculated using nine orbitals, which can be labeled as  $\sigma_g$, $\sigma_g$, $\sigma_u$, $\sigma_u$ ,$\pi_{u,\pm 1}$, $\sigma_g$, $\pi_{g, \pm 1}$  at the beginning of the propagation. These calculations have a spin adapted triplet configuration space of dimension 36.  We used a fixed nuclei Hamiltonian where the internuclear distance is 2.282 bohr. Prolate spheroidal coordinates, $(\eta,\xi,\phi)$, were used in these calculations, and we employ a DVR grid of $10$ points for $\eta$ and a $\xi$ grid with twelve grid points per finite element.  Nine finite elements were used in $\xi$, the first of length 2.0 a$_0$ providing a dense grid to represent the 1s orbital and orbital cusp region, with seven subsequent five elements  of length 8.0 a$_0$. Exterior complex scaling is applied to the remaining four elements extending an additional 32 a$_0$ with a complex scaling angle of 0.40 radians.  The results show no sensitivity to the ECS angle, indicating that all ionized flux is being completely absorbed by the ECS procedure. 

The MCTDHF calculation describes the relative energies and line shapes in these cross section, but does not reproduce the absolute excitation energies of these autoionizing states.  Thus, the calculated results were shifted to lower energies by $5.18$ eV such that the limit of the calculated Rydberg series, corresponds to the c$^4\Sigma_u^-$ state of O$_2^+$, agrees with the literature value. Because the present calculations used a fixed nuclei treatment, the MCTDHF computed cross section does not exhibit vibrational structure.  We observe that the relative energies of the (2$\sigma_u$)$^{-1}$(ns$\sigma_g$) and (2$\sigma_u$)$^{-1}$(nd$\sigma_g$) series from our fixed-nuclei MCTDHF calculations agree well with the locations of first vibrational states ($\nu$=0) of those two series as in ~\cite{Holland, Demekhin}.

The MCTDHF computed cross section is for randomly oriented molecules in the presence of a linearly polarized XUV field. This total cross section  is computed using the appropriate relation for single photon absorption, $ \sigma_{\mbox{total}}=\frac{1}{3}\sigma_{ \parallel}+\frac{2}{3}\sigma_{\perp}$, where $\sigma_{\parallel}$ and $\sigma_{\perp}$ are calculated separately for oriented molecules either parallel or perpendicular to the polarization directions of the fields.  The calculations from Demekhin {\it et al.}~\cite{Demekhin}, considered only parallel polarization,  and thus there were only two Rydberg series, (2$\sigma_u$)$^{-1}$(ns$\sigma_g$) and (2$\sigma_u$)$^{-1}$(nd$\sigma_g$), converging to the c$^4\Sigma_u^-$ limit. However, using perpendicular polarization, the MCTDHF computed $\sigma_{\perp}$ exhibits a third Rydberg series converging to the same limit, namely, the series we have identified as (2$\sigma_u$)$^{-1}$(nd$\pi_g$) here.   

\begin{figure}[htb!]
	\includegraphics[width=0.95\columnwidth,clip=true]{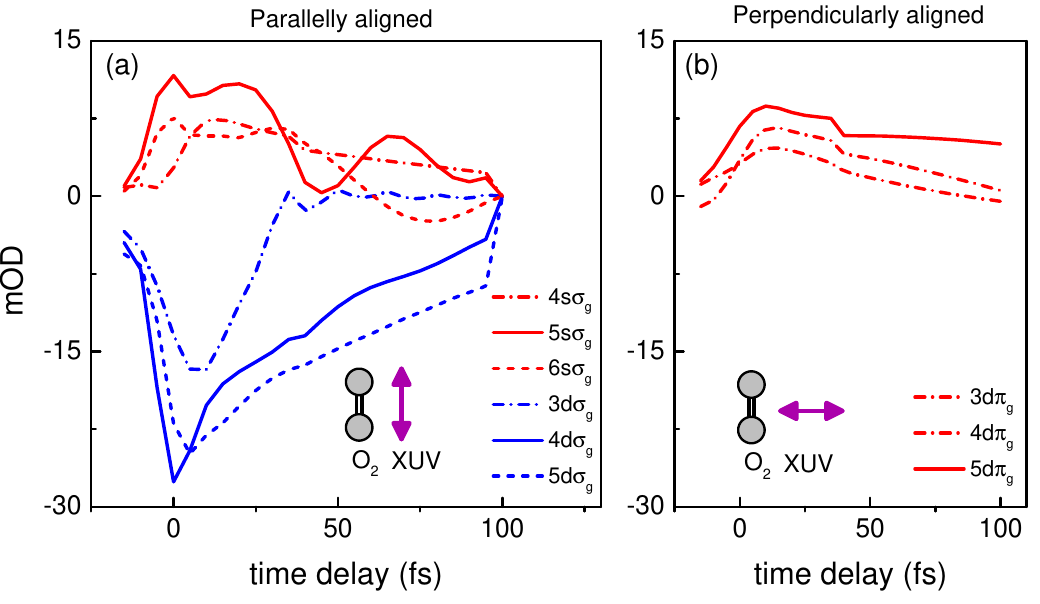}
	\caption{
		(Color online) ~\Liao{Transient absorption line-outs calculated using MCTDHF method for molecules (a)  aligned parallel to XUV field polarization, and (b) perpendicularly to XUV field polarization, plotted at the energies corresponding to several resonances. Note that the NIR field polarization is always parallel to the XUV field polarization.}}
	\label{FigPerPara} 
\end{figure}

To compute quantities directly comparable to the experimental observation of the quantity in Eq.(\ref{eq:OD}), we begin with Beer-Lambert law, $-\log [I_{\mbox{out}}/I_{\mbox{in}}]=\sigma \, N \, L $ where $I_{\mbox{in}}$ and $I_{\mbox{out}}$ are the incoming and outgoing field intensities, respectively, and $\sigma$ is the photoabsorption cross section. For these comparisons, we estimated the molecular density as $N=1 \times 10^{16}$cm$^{-3}$  and $L=1$ cm for the path length. The photoabsorption cross section is related to the response functions, $\tilde{S}(\omega)$, computed using the MCTDHF method. Therefore, we can construct the appropriate OD corresponding to Eq.(\ref{eq:OD})  as
\begin{equation}
\begin{split}
\mbox{OD} = & -\log \Big[\frac{I_{\mbox{\tiny out}}^{\scalebox{.5}{XUV+NIR}}(t_d)}{I_{\mbox{\tiny out}}^{\scalebox{.5}{XUV}} }\Big] \\
= &- \left\{ \log \Big[ \frac{I_{\mbox{\tiny out}}^{\scalebox{.5}{XUV+NIR}}(t_d)}{I_{\mbox{ \tiny in}}^{\scalebox{.5}{XUV+NIR}}}\Big] -\log \Big[\frac{I_{\mbox{\tiny out}}^{\scalebox{.5}{XUV}}}{I_{\mbox{\tiny in}}^{\scalebox{.5}{XUV}}}\Big] \right\}  \\
= & 4\pi\alpha\omega \bigg[\frac{\tilde{S}_{\scalebox{.5}{XUV+NIR}}(\omega;t_d)}{|\tilde{\mathcal{E}}_{\mbox{\tiny in}}^{\scalebox{.5}{XUV+NIR}}(\omega)|^2} -\frac{\tilde{S}_{\scalebox{.5}{XUV}}(\omega)}{|\tilde{\mathcal{E}}_{\mbox{\tiny in}}^{\scalebox{.5}{XUV}}(\omega)|^2}\bigg]NL \\
\end{split}
\label{eq:OD0}
\end{equation}

At frequencies in the XUV,  the contribution of the NIR pulse is negligible, so that $\tilde{\mathcal{E}}_{\mbox{\tiny in}}^{\scalebox{.5}{XUV+NIR}}(\omega) \approx \tilde{\mathcal{E}}_{\mbox{\tiny in}}^{\scalebox{.5}{XUV}}(\omega)$. ~\Liao{With} that assumption, Eq.(\ref{eq:OD0})  becomes the following working expression for the measured quantity in terms of the calculated frequency- and delay-dependent response functions,
\begin{eqnarray}
\label{eq:opticaldensity}
\mbox{OD}\approx 4\pi\alpha\omega \frac{\Big[\tilde{S}_{\scalebox{.5}{XUV+NIR}}(\omega;t_d)-\tilde{S}_{\scalebox{.5}{XUV}}(\omega)\Big]}{|\tilde{\mathcal{E}}_{\mbox{\tiny in}}^{\scalebox{.5}{XUV}}(\omega)|^2}NL.
\end{eqnarray}

\Liao{When the molecular axis of O$_2$ is parallel to both XUV and NIR field polarization directions as in Fig.~\ref{FigPerPara} (a), the nd$\sigma_g$ states exhibit negative OD, while ns$\sigma_g$ states  show positive OD. However, as shown in Fig.~\ref{FigPerPara} (b), where the molecular axis is perpendicular to the polarization directions of both XUV and NIR fields, the nd$\pi_g$ states contribute positive OD.} In order to compute the time-delay dependent OD for randomly oriented molecules, we make the approximation (exact for one-photon absorption), $\mbox{OD}_{\mbox{total}}=\frac{1}{3}\mbox{OD}_{\parallel}+\frac{2}{3}\mbox{OD}_{\perp}$, and the result is shown in Fig.~\ref{FigATAspectra} (b) and Fig.~\ref{FigSpectrogram} (d).  ~\Liao{Specifically, in Fig.~\ref{FigSpectrogram} (d) we show OD at six energies which correspond to three resonances for the (2$\sigma_u$)$^{-1}$nd$\sigma_g$  series, and three resonances for the (2$\sigma_u$)$^{-1}$(nd$\pi_g$)+(2$\sigma_u$)$^{-1}$(ns$\sigma_g$) series.} 

\section{Few-level Model for Transient Absorption}
\label{sec:FewLevelModel}

\subsection{XUV initiated polarization}

We first substitute Eq.(\ref{eq:total}) and the Hamiltonian in Eq.(\ref{Hamiltonian}) into the Schr\"odinger equation, and then project onto that equation with $\langle g |$, $\langle b |$, and $\langle E |$.  We then assume that the only nonzero dipole matrix elements are
$D_{gb}= \langle g|\hat{d}|b\rangle$ and $D_{gE} = \langle g|\hat{d}|E \rangle$, and recalling that the XUV field is $\mathcal{E}_{\tiny \mbox{XUV}}(t)=F(t) e^{i\omega_{\scalebox{.5}{XUV}} t}+c.c.$, make the rotating wave approximation. ~\Liao{We also} assume that the molecular Hamiltonian, $H_0$, couples only $|b\rangle$ to $| E \rangle$, so that its only nonzero matrix elements are $E_g = \langle g|H_0|g \rangle$, $E_b = \langle b|H_0|b \rangle$, $\langle E |  H_0 | E' \rangle = \delta(E-E')$, and $V_{bE} = \langle b | H_0 | E \rangle$. Then, using the orthogonality of $|g\rangle$, $|b \rangle$, and $|E \rangle$, we find that the time-dependent coefficients $C_g(t)$, $C_b(t)$, and $C_E(t)$ satisfy the coupled differential equations,

\begin{eqnarray}{\label{eq:coupled}}
i\dot{C}_g&=&-\left[\int dE~D_{gE}C_EF(t)+D_{gb}C_bF(t)\right],\\
i\dot{C}_b&=&[E_b-(E_g+\omega_{\scalebox{.5}{XUV}})]C_b\nonumber\\
&+&\int dE~V_{bE}C_E-D_{gb}^*C_g {F(t)^*},\\
i\dot{C}_E&=&[E-(E_g+\omega_{\scalebox{.5}{XUV}})]C_E\nonumber\\
&+&V_{Eb}C_b-D_{Eg}C_g {F(t)^*}.
\end{eqnarray}

We also adopt the adiabatic elimination of the continuum by assuming that the coefficients of the continuum states change much more slowly than those of the discrete states,  i.e., $\dot{C}_E \approx 0$, as was done in Ref.~\cite{CDLin2013}. The equation for the time-dependent coefficients for the continuum states,  $C_E(t)$,  can thus be simplified to give 
\begin{equation}{\label{eq:CE}}
C_E=-\frac{V_{Eb}C_b-D_{Eg}C_g F(t)^* }{E-(E_g+\omega_{\scalebox{.5}{XUV}})}.
\end{equation}
The coefficient $C_E(t)$ appears in the coupled equations under the integration $\int dE$ over continuum states, so this equation is actually a representation of a Green's function, namely, 
\begin{equation}{\label{{eq:CE2}}}
C_E=\lim_{\epsilon \rightarrow 0^+} \frac{V_{Eb}C_b-D_{Eg}C_g {F(t)^*} }{(E_g+\omega_{\tiny \mbox{XUV}})-E + i \epsilon}.
\end{equation}

We then retain only the resonant contribution to integral over $E$ in the Eq.(\ref{eq:coupled}) and the next equation, i.e., the contribution ~\Liao{is} proportional to $-i \pi \delta(E_g + \omega_{\scalebox{.5}{XUV}} - E)$. Making use of the definitions of Fano $q$ parameter, we arrive at simplified equations for $C_g(t)$ and $C_b(t)$ for the discrete states,
\begin{equation}{\label{{eq:Cg}}}
i\dot{C}_g=-i\pi |D_{gE}|^2 |F(t)|^2C_g-D_{gb}C_bF(t)(1-i/q)
\end{equation}
\begin{equation}{\label{{eq:iCb}}}
i\dot{C}_b=[\Delta_b-i\Gamma/2]C_b-D_{gb}^*C_g F(t)^* (1-i/q),
\end{equation}
with $D_{gE}$ evaluated at $E = E_g + \omega_{\tiny \mbox{XUV}}$ and $\Delta_b = E_b -(E_g+\omega_{\tiny \mbox{XUV}})$. Here, $q$ and $\Gamma$ are the conventionally defined Fano $q$ parameters that describe the line shape and the width,
\begin{eqnarray}
q = \frac{D_{gb}}{\pi D_{gE}V_{Eb}}\quad , \\
 \quad \Gamma =2 \pi |V_{Eb}|^2 \quad .
\label{eq:Fano}
\end{eqnarray}
Going back to the original expression for the $\Psi(t)$ in  Eq.(\ref{eq:total}), we can express the  leading contribution to the time-dependent dipole as
\begin{equation}
\begin{split}
d(t) \equiv & \langle \Psi(t)| \hat d | \Psi(t) \rangle\\
=&\exp{(-i\omega_{\scalebox{.5}{XUV}} t)}C_g^* \left[D_{gb}C_b+\int dE~D_{gE}C_E \right]+\mbox{c.c.} \\
=&\exp{(-i\omega_{\scalebox{.5}{XUV}} t)} \left[C_g^*D_{gb}C_b(1-i/q) \right.  \\
&\left.  \qquad  \qquad + i\pi|C_g|^2 |D_{gE}|^2 F(t)^* \right]+\mbox{c.c.}\\
\end{split}
\end{equation}
So we need only $C_g(t)$ and $C_b(t)$ to evaluate this expression.  We now make the approximation that the ground state is not appreciably depopulated, $C_g(t) \approx 1$, note also that on resonance $\Delta_b = 0$, and solve the equation for $C_b$ using the substitution $C_b = \exp({-\, \Gamma t /2} )\, \bar{C}_b $ in the case of the Gaussian XUV pulse in Eq.(\ref{eq:GaussPulse}) to obtain
\begin{equation}
C_b(t) = i \,  e^{-{\Gamma t}/{2}}  D_{gb}^* (1-i/q) F_{\scalebox{.5} {XUV}} G(t), 
\end{equation}
where ~\Liao{we define}
\begin{equation}
\begin{split}
G(t) \equiv  & ~\frac{1}{2} \,  e^{\Gamma^2 \tau_{\scalebox{.5}{XUV}}^2 /{16}  } \bigg[1 
+ \erf\left({\Gamma\tau_{\scalebox{.5} {XUV}}}/{4} + {t}/{\tau_{\scalebox{.5} {XUV}}}\right)\bigg], 
\end{split}
\end{equation}
and $\erf$ is the error function. Assembling $d(t)$ we then find
\begin{equation}{\label{eq:DipoleExpression}}
\begin{split}
d(t)=
 \quad & e^{-i \omega_{\scalebox{.5}{XUV}} t} \bigg[i |D_{gb}|^2 
(1-i/q)^2 F_{\scalebox{.5}{XUV}} \,  G(t) \, e^{-\Gamma t/{2} } \\
&+ i \pi |D_{gE}|^2 F_{\scalebox{.5}{XUV}} \, e^{-t^2/\tau_{\scalebox{.5} {XUV}}^2} \left(\pi \tau_{\scalebox{.5} {XUV}}^2\right)^{-1/2}  \bigg] 
+\mbox{c.c.}.
\end{split}
\end{equation}
This is the expression for the polarization that we use to model transient absorption, by modifying it with either laser-induced attenuation or laser-induced phase shift as described in Sec. ~\ref{sec:Model}, and then Fourier transforming it according to 
\begin{equation}
\tilde d(\omega) = \frac{1}{\sqrt{2 \pi}} \int^\infty_{- \infty}  dt \, e^{i \omega t} d(t) 
\end{equation}
to construct the response function in Eq.(\ref{eq:ResponseFunc}) that gives the transient absorption spectrum.

To make the connection with Fano line shapes in the pure XUV absorption spectrum we evaluate Eq.(\ref{eq:DipoleExpression}) in the limit of a delta function XUV pulse, $\tau_{\scalebox{.5} {XUV}} \rightarrow 0$~\Liao{, so that}
\begin{equation}
d(t) = 
\left\{
\begin{array}{cc}
0 & ,t<0 \\
e^{-i \, \omega_{\tiny \mbox{XUV}} t}  \left[ i \left| D_{gb} \right|^2  (1-i/q)^2 F_{\scalebox{.5}{XUV}} e^{-\frac{\Gamma}{2} t}  \right. \\ 
\qquad \qquad + \left. i \pi \left| D_{gE} \right|^2 F_{\scalebox{.5}{XUV}} \delta(t) \right]  & ,t \ge 0
\end{array}
\right. ,
\label{eq:doft}
\end{equation}
and the Fourier transform of the ~\Liao{polarization} becomes 
\begin{equation}
\tilde{d}(\omega)=\frac{1}{\sqrt{2\pi}}F_{\scalebox{.5}{XUV}} \left[i\pi|D_{gE}|^2  -   \frac{(1-i/q)^2|D_{gb}|^2}{\omega-\omega_{\tiny \mbox{XUV}} + i \, \Gamma/2}\right].
\end{equation}

With this approximation to $\tilde{d}(\omega)$, and the Fourier transform of the corresponding XUV field, $\tilde{\mathcal{E}}_{\tiny \mbox{XUV} } (\omega) = F_{\scalebox{.5}{XUV}} /\sqrt{2\pi}  $, the ~\Liao{absorption cross section now becomes}
\begin{eqnarray}
\sigma(\omega)&=&8\pi\alpha\omega \, \mbox{Im}\left[ \tilde{d}(\omega)/ \mathcal{E}(\omega)\right]\nonumber\\
&=&8\pi^2\alpha\, \omega |D_{gE}|^2 \frac{(\omega-\omega_{\tiny \mbox{XUV}}+q \, \Gamma/2)^2}{(\omega-\omega_{\tiny \mbox{XUV}})^2+(\Gamma/2)^2},
\label{eq:FanoProfile}
\end{eqnarray}
which has the well known form of the absorption cross section in the vicinity of an autoionizing feature with a Fano resonance line shape.   In this limit of a delta function XUV pulse, it is possible to analytically evaluate the Fourier transform of the modified d(t)  in the case of sudden attenuation in the LIA model in Eq.(\ref{eq:LIA}) and see explicitly how it depends on the Fano $q$ parameters of the unperturbed line shape. The change in the response function due to this sudden truncation of the polarization is
\begin{equation}
\Delta \tilde{S}(\omega;t_d)\equiv  \tilde{S}(\omega;t_d)-\tilde{S}_{\tiny \mbox{XUV}}(\omega),
\end{equation}
and the final expression for $\Delta \tilde{S}(\omega;t_d)$ in this simple model, evaluated at the resonance energy $\omega =\omega_{\tiny \mbox{XUV}}$ can then be shown to be 
\begin{equation}
{\label{eq:suddenLIA}}
 \Delta \tilde{S}(\omega=\omega_{\tiny \mbox{XUV}};t_d)=  |F_0|^2 |D_{gE}|^2 (1-q^2) e^{-\Gamma t_d/2}.
\end{equation}

\Liao{As we can see that} in this extreme version of the LIA model, the dependence on Fano $q$ parameter of the change in the response function is negative or positive depending on whether $q$ is greater or less than unity because of the factor of $1-q^2$ in Eq.(\ref{eq:suddenLIA}).

\subsection{NIR perturbation: The LIA and LIP models}

We can extend the above approach to set up two superimposed polarizations corresponding to ~\Liao{a pair of features of nd$\sigma_g$ and ns$\sigma_g$+nd$\pi_g$, as the weighted sum of two polarizations $C_{nd} \ d^{\tiny nd}(t;q_{nd},\Gamma_{nd}) + C_{ns} \ d^{\tiny ns}(t;q_{ns},\Gamma_{ns})$. The polarization labeled $d^{\tiny ns}$ includes ns$\sigma_g$+nd$\pi_g$ contributions, and the weighting coefficients $C_{nd}$ and $C_{ns}$ are chosen to reproduce static absorption spectrum as in our Schwinger calculation.}

We then calculate the response function as defined in Eq.(\ref{eq:ResponseFunc}) in spectral domain. The parameters, such as Fano $q$, the field-free decay life times $\Gamma$, and the amplitudes of dipole matrix elements $D_{gb}$ and $D_{bE}$, are chosen so that the response function closely matches the static absorption line shapes shown in Fig.~\ref{FigPEcurves}(b). The delay-dependent NIR perturbations are then modeled as LIA or LIP. In the LIA method, XUV initiated polarizations are attenuated slowly by an error function profile centered at some delay $t_d$ to simulate the effect of excited state population removal by Gaussian shaped NIR pulse. The parameters used for the LIP calculations that determine the pondermotive energy function ~\Liao{were $\mathcal{E}^{0}_{\scalebox{.5}{NIR}}$  = 5.3 $\times$ 10$^{-3}$ a.u., corresponding to an intensity of 1 TW/cm$^2$, and the NIR photon energy $\omega_L$ = 0.058 a.u. (1.58 eV)}.

\bibliography{O2paper_ref_v7}

\begin{thebibliography}{68}%
\makeatletter
\providecommand \@ifxundefined [1]{%
 \@ifx{#1\undefined}
}%
\providecommand \@ifnum [1]{%
 \ifnum #1\expandafter \@firstoftwo
 \else \expandafter \@secondoftwo
 \fi
}%
\providecommand \@ifx [1]{%
 \ifx #1\expandafter \@firstoftwo
 \else \expandafter \@secondoftwo
 \fi
}%
\providecommand \natexlab [1]{#1}%
\providecommand \enquote  [1]{``#1''}%
\providecommand \bibnamefont  [1]{#1}%
\providecommand \bibfnamefont [1]{#1}%
\providecommand \citenamefont [1]{#1}%
\providecommand \href@noop [0]{\@secondoftwo}%
\providecommand \href [0]{\begingroup \@sanitize@url \@href}%
\providecommand \@href[1]{\@@startlink{#1}\@@href}%
\providecommand \@@href[1]{\endgroup#1\@@endlink}%
\providecommand \@sanitize@url [0]{\catcode `\\12\catcode `\$12\catcode
  `\&12\catcode `\#12\catcode `\^12\catcode `\_12\catcode `\%12\relax}%
\providecommand \@@startlink[1]{}%
\providecommand \@@endlink[0]{}%
\providecommand \url  [0]{\begingroup\@sanitize@url \@url }%
\providecommand \@url [1]{\endgroup\@href {#1}{\urlprefix }}%
\providecommand \urlprefix  [0]{URL }%
\providecommand \Eprint [0]{\href }%
\providecommand \doibase [0]{http://dx.doi.org/}%
\providecommand \selectlanguage [0]{\@gobble}%
\providecommand \bibinfo  [0]{\@secondoftwo}%
\providecommand \bibfield  [0]{\@secondoftwo}%
\providecommand \translation [1]{[#1]}%
\providecommand \BibitemOpen [0]{}%
\providecommand \bibitemStop [0]{}%
\providecommand \bibitemNoStop [0]{.\EOS\space}%
\providecommand \EOS [0]{\spacefactor3000\relax}%
\providecommand \BibitemShut  [1]{\csname bibitem#1\endcsname}%
\let\auto@bib@innerbib\@empty
\bibitem [{\citenamefont {Becker}\ and\ \citenamefont
  {Shirley}(2012)}]{becker2012}%
  \BibitemOpen
  \bibfield  {author} {\bibinfo {author} {\bibfnamefont {U.}~\bibnamefont
  {Becker}}\ and\ \bibinfo {author} {\bibfnamefont {D.~A.}\ \bibnamefont
  {Shirley}},\ }\href@noop {} {\emph {\bibinfo {title} {VUV and Soft X-ray
  Photoionization}}}\ (\bibinfo  {publisher} {Springer Science \& Business
  Media},\ \bibinfo {year} {2012})\BibitemShut {NoStop}%
\bibitem [{\citenamefont {Ng}(1991)}]{ng1991}%
  \BibitemOpen
  \bibfield  {author} {\bibinfo {author} {\bibfnamefont {C.-Y.}\ \bibnamefont
  {Ng}},\ }\href@noop {} {\emph {\bibinfo {title} {Vacuum ultraviolet
  photoionization and photodissociation of molecules and clusters}}}\ (\bibinfo
   {publisher} {World Scientific},\ \bibinfo {year} {1991})\BibitemShut
  {NoStop}%
\bibitem [{\citenamefont {Platzman}(1962)}]{platzman1962superexcited}%
  \BibitemOpen
  \bibfield  {author} {\bibinfo {author} {\bibfnamefont {R.~L.}\ \bibnamefont
  {Platzman}},\ }\href@noop {} {\bibfield  {journal} {\bibinfo  {journal}
  {Radiation Research}\ }\textbf {\bibinfo {volume} {17}},\ \bibinfo {pages}
  {419} (\bibinfo {year} {1962})}\BibitemShut {NoStop}%
\bibitem [{\citenamefont {Hatano}(1999)}]{Hatano1999}%
  \BibitemOpen
  \bibfield  {author} {\bibinfo {author} {\bibfnamefont {Y.}~\bibnamefont
  {Hatano}},\ }\href@noop {} {\bibfield  {journal} {\bibinfo  {journal}
  {Physics Reports-Review Section of Physics Letters}\ }\textbf {\bibinfo
  {volume} {313}},\ \bibinfo {pages} {110} (\bibinfo {year}
  {1999})}\BibitemShut {NoStop}%
\bibitem [{\citenamefont {Nakamura}(1991)}]{nakamura1991basic}%
  \BibitemOpen
  \bibfield  {author} {\bibinfo {author} {\bibfnamefont {H.}~\bibnamefont
  {Nakamura}},\ }\href@noop {} {\bibfield  {journal} {\bibinfo  {journal}
  {International Reviews in Physical Chemistry}\ }\textbf {\bibinfo {volume}
  {10}},\ \bibinfo {pages} {123} (\bibinfo {year} {1991})}\BibitemShut
  {NoStop}%
\bibitem [{\citenamefont {Wayne}(1991)}]{Wayne1991}%
  \BibitemOpen
  \bibfield  {author} {\bibinfo {author} {\bibfnamefont {R.~P.}\ \bibnamefont
  {Wayne}},\ }\href@noop {} {\emph {\bibinfo {title} {Chemistry of Atmosphere:
  An Introduction to the Chemistry of the Atmosphere of Earth, the Planets and
  Their Satellites}}}\ (\bibinfo  {publisher} {Oxford},\ \bibinfo {address}
  {Clarendon},\ \bibinfo {year} {1991})\BibitemShut {NoStop}%
\bibitem [{\citenamefont {Bouda{\i}ffa}\ \emph {et~al.}(2000)\citenamefont
  {Bouda{\i}ffa}, \citenamefont {Cloutier}, \citenamefont {Hunting},
  \citenamefont {Huels},\ and\ \citenamefont {Sanche}}]{Bouda.2000.DNAdamage}%
  \BibitemOpen
  \bibfield  {author} {\bibinfo {author} {\bibfnamefont {B.}~\bibnamefont
  {Bouda{\i}ffa}}, \bibinfo {author} {\bibfnamefont {P.}~\bibnamefont
  {Cloutier}}, \bibinfo {author} {\bibfnamefont {D.}~\bibnamefont {Hunting}},
  \bibinfo {author} {\bibfnamefont {M.~A.}\ \bibnamefont {Huels}}, \ and\
  \bibinfo {author} {\bibfnamefont {L.}~\bibnamefont {Sanche}},\ }\href@noop {}
  {\bibfield  {journal} {\bibinfo  {journal} {Science}\ }\textbf {\bibinfo
  {volume} {287}},\ \bibinfo {pages} {1658} (\bibinfo {year}
  {2000})}\BibitemShut {NoStop}%
\bibitem [{\citenamefont {Florescu-Mitchell}\ and\ \citenamefont
  {Mitchell}(2006)}]{FlorescuMitchell2006}%
  \BibitemOpen
  \bibfield  {author} {\bibinfo {author} {\bibfnamefont {A.}~\bibnamefont
  {Florescu-Mitchell}}\ and\ \bibinfo {author} {\bibfnamefont {J.}~\bibnamefont
  {Mitchell}},\ }\href {\doibase
  http://dx.doi.org/10.1016/j.physrep.2006.04.002} {\bibfield  {journal}
  {\bibinfo  {journal} {Physics Reports}\ }\textbf {\bibinfo {volume} {430}},\
  \bibinfo {pages} {277 } (\bibinfo {year} {2006})}\BibitemShut {NoStop}%
\bibitem [{\citenamefont {Kokoouline}\ \emph {et~al.}(2011)\citenamefont
  {Kokoouline}, \citenamefont {Douguet},\ and\ \citenamefont
  {Greene}}]{KokooulineGreene2011}%
  \BibitemOpen
  \bibfield  {author} {\bibinfo {author} {\bibfnamefont {V.}~\bibnamefont
  {Kokoouline}}, \bibinfo {author} {\bibfnamefont {N.}~\bibnamefont {Douguet}},
  \ and\ \bibinfo {author} {\bibfnamefont {C.~H.}\ \bibnamefont {Greene}},\
  }\href {\doibase http://dx.doi.org/10.1016/j.cplett.2011.03.062} {\bibfield
  {journal} {\bibinfo  {journal} {Chemical Physics Letters}\ }\textbf {\bibinfo
  {volume} {507}},\ \bibinfo {pages} {1 } (\bibinfo {year} {2011})}\BibitemShut
  {NoStop}%
\bibitem [{\citenamefont {Douguet}\ \emph {et~al.}(2012)\citenamefont
  {Douguet}, \citenamefont {Orel}, \citenamefont {Greene},\ and\ \citenamefont
  {Kokoouline}}]{Douguet_Orel_Greene2012}%
  \BibitemOpen
  \bibfield  {author} {\bibinfo {author} {\bibfnamefont {N.}~\bibnamefont
  {Douguet}}, \bibinfo {author} {\bibfnamefont {A.~E.}\ \bibnamefont {Orel}},
  \bibinfo {author} {\bibfnamefont {C.~H.}\ \bibnamefont {Greene}}, \ and\
  \bibinfo {author} {\bibfnamefont {V.}~\bibnamefont {Kokoouline}},\ }\href
  {\doibase 10.1103/PhysRevLett.108.023202} {\bibfield  {journal} {\bibinfo
  {journal} {Phys. Rev. Lett.}\ }\textbf {\bibinfo {volume} {108}},\ \bibinfo
  {pages} {023202} (\bibinfo {year} {2012})}\BibitemShut {NoStop}%
\bibitem [{\citenamefont {Kokoouline}\ \emph {et~al.}(2001)\citenamefont
  {Kokoouline}, \citenamefont {Greene},\ and\ \citenamefont
  {Esry}}]{Kokoouline_Greene_Esry_2001}%
  \BibitemOpen
  \bibfield  {author} {\bibinfo {author} {\bibfnamefont {V.}~\bibnamefont
  {Kokoouline}}, \bibinfo {author} {\bibfnamefont {C.~H.}\ \bibnamefont
  {Greene}}, \ and\ \bibinfo {author} {\bibfnamefont {B.~D.}\ \bibnamefont
  {Esry}},\ }\href@noop {} {\bibfield  {journal} {\bibinfo  {journal} {Nature}\
  }\textbf {\bibinfo {volume} {412}},\ \bibinfo {pages} {891} (\bibinfo {year}
  {2001})}\BibitemShut {NoStop}%
\bibitem [{\citenamefont {Guberman}\ and\ \citenamefont
  {Giusti-Suzor}(1991)}]{Guberman1991}%
  \BibitemOpen
  \bibfield  {author} {\bibinfo {author} {\bibfnamefont {S.~L.}\ \bibnamefont
  {Guberman}}\ and\ \bibinfo {author} {\bibfnamefont {A.}~\bibnamefont
  {Giusti-Suzor}},\ }\href {\doibase http://dx.doi.org/10.1063/1.460913}
  {\bibfield  {journal} {\bibinfo  {journal} {The Journal of Chemical Physics}\
  }\textbf {\bibinfo {volume} {95}},\ \bibinfo {pages} {2602} (\bibinfo {year}
  {1991})}\BibitemShut {NoStop}%
\bibitem [{\citenamefont {Jungen}\ and\ \citenamefont
  {Pratt}(2010)}]{Jungen_Pratt2010}%
  \BibitemOpen
  \bibfield  {author} {\bibinfo {author} {\bibfnamefont {C.}~\bibnamefont
  {Jungen}}\ and\ \bibinfo {author} {\bibfnamefont {S.~T.}\ \bibnamefont
  {Pratt}},\ }\href@noop {} {\bibfield  {journal} {\bibinfo  {journal} {The
  Journal of Chemical Physics}\ }\textbf {\bibinfo {volume} {133}},\ \bibinfo
  {pages} {214303} (\bibinfo {year} {2010})}\BibitemShut {NoStop}%
\bibitem [{\citenamefont {Rundquist}\ \emph {et~al.}(1998)\citenamefont
  {Rundquist}, \citenamefont {Durfee}, \citenamefont {Chang}, \citenamefont
  {Herne}, \citenamefont {Backus}, \citenamefont {Murnane},\ and\ \citenamefont
  {Kapteyn}}]{rundquist.1998.HHG}%
  \BibitemOpen
  \bibfield  {author} {\bibinfo {author} {\bibfnamefont {A.}~\bibnamefont
  {Rundquist}}, \bibinfo {author} {\bibfnamefont {C.~G.}\ \bibnamefont
  {Durfee}}, \bibinfo {author} {\bibfnamefont {Z.}~\bibnamefont {Chang}},
  \bibinfo {author} {\bibfnamefont {C.}~\bibnamefont {Herne}}, \bibinfo
  {author} {\bibfnamefont {S.}~\bibnamefont {Backus}}, \bibinfo {author}
  {\bibfnamefont {M.~M.}\ \bibnamefont {Murnane}}, \ and\ \bibinfo {author}
  {\bibfnamefont {H.~C.}\ \bibnamefont {Kapteyn}},\ }\href@noop {} {\bibfield
  {journal} {\bibinfo  {journal} {Science}\ }\textbf {\bibinfo {volume}
  {280}},\ \bibinfo {pages} {1412} (\bibinfo {year} {1998})}\BibitemShut
  {NoStop}%
\bibitem [{\citenamefont {Paul}\ \emph {et~al.}(2001)\citenamefont {Paul},
  \citenamefont {Toma}, \citenamefont {Breger}, \citenamefont {Mullot},
  \citenamefont {Aug{\'e}}, \citenamefont {Balcou}, \citenamefont {Muller},\
  and\ \citenamefont {Agostini}}]{paul.2001.APTs}%
  \BibitemOpen
  \bibfield  {author} {\bibinfo {author} {\bibfnamefont {P.}~\bibnamefont
  {Paul}}, \bibinfo {author} {\bibfnamefont {E.}~\bibnamefont {Toma}}, \bibinfo
  {author} {\bibfnamefont {P.}~\bibnamefont {Breger}}, \bibinfo {author}
  {\bibfnamefont {G.}~\bibnamefont {Mullot}}, \bibinfo {author} {\bibfnamefont
  {F.}~\bibnamefont {Aug{\'e}}}, \bibinfo {author} {\bibfnamefont
  {P.}~\bibnamefont {Balcou}}, \bibinfo {author} {\bibfnamefont
  {H.}~\bibnamefont {Muller}}, \ and\ \bibinfo {author} {\bibfnamefont
  {P.}~\bibnamefont {Agostini}},\ }\href@noop {} {\bibfield  {journal}
  {\bibinfo  {journal} {Science}\ }\textbf {\bibinfo {volume} {292}},\ \bibinfo
  {pages} {1689} (\bibinfo {year} {2001})}\BibitemShut {NoStop}%
\bibitem [{\citenamefont {Gagnon}\ \emph {et~al.}(2007)\citenamefont {Gagnon},
  \citenamefont {Ranitovic}, \citenamefont {Tong}, \citenamefont {Cocke},
  \citenamefont {Murnane}, \citenamefont {Kapteyn},\ and\ \citenamefont
  {Sandhu}}]{gagnon.2007.MO}%
  \BibitemOpen
  \bibfield  {author} {\bibinfo {author} {\bibfnamefont {E.}~\bibnamefont
  {Gagnon}}, \bibinfo {author} {\bibfnamefont {P.}~\bibnamefont {Ranitovic}},
  \bibinfo {author} {\bibfnamefont {X.-M.}\ \bibnamefont {Tong}}, \bibinfo
  {author} {\bibfnamefont {C.~L.}\ \bibnamefont {Cocke}}, \bibinfo {author}
  {\bibfnamefont {M.~M.}\ \bibnamefont {Murnane}}, \bibinfo {author}
  {\bibfnamefont {H.~C.}\ \bibnamefont {Kapteyn}}, \ and\ \bibinfo {author}
  {\bibfnamefont {A.~S.}\ \bibnamefont {Sandhu}},\ }\href@noop {} {\bibfield
  {journal} {\bibinfo  {journal} {Science}\ }\textbf {\bibinfo {volume}
  {317}},\ \bibinfo {pages} {1374} (\bibinfo {year} {2007})}\BibitemShut
  {NoStop}%
\bibitem [{\citenamefont {Sandhu}\ \emph {et~al.}(2008)\citenamefont {Sandhu},
  \citenamefont {Gagnon}, \citenamefont {Santra}, \citenamefont {Sharma},
  \citenamefont {Li}, \citenamefont {Ho}, \citenamefont {Ranitovic},
  \citenamefont {Cocke}, \citenamefont {Murnane},\ and\ \citenamefont
  {Kapteyn}}]{sandhu2008.O2}%
  \BibitemOpen
  \bibfield  {author} {\bibinfo {author} {\bibfnamefont {A.~S.}\ \bibnamefont
  {Sandhu}}, \bibinfo {author} {\bibfnamefont {E.}~\bibnamefont {Gagnon}},
  \bibinfo {author} {\bibfnamefont {R.}~\bibnamefont {Santra}}, \bibinfo
  {author} {\bibfnamefont {V.}~\bibnamefont {Sharma}}, \bibinfo {author}
  {\bibfnamefont {W.}~\bibnamefont {Li}}, \bibinfo {author} {\bibfnamefont
  {P.}~\bibnamefont {Ho}}, \bibinfo {author} {\bibfnamefont {P.}~\bibnamefont
  {Ranitovic}}, \bibinfo {author} {\bibfnamefont {C.~L.}\ \bibnamefont
  {Cocke}}, \bibinfo {author} {\bibfnamefont {M.~M.}\ \bibnamefont {Murnane}},
  \ and\ \bibinfo {author} {\bibfnamefont {H.~C.}\ \bibnamefont {Kapteyn}},\
  }\href@noop {} {\bibfield  {journal} {\bibinfo  {journal} {Science}\ }\textbf
  {\bibinfo {volume} {322}},\ \bibinfo {pages} {1081} (\bibinfo {year}
  {2008})}\BibitemShut {NoStop}%
\bibitem [{\citenamefont {Krausz}\ and\ \citenamefont
  {Ivanov}(2009)}]{krausz2009}%
  \BibitemOpen
  \bibfield  {author} {\bibinfo {author} {\bibfnamefont {F.}~\bibnamefont
  {Krausz}}\ and\ \bibinfo {author} {\bibfnamefont {M.}~\bibnamefont
  {Ivanov}},\ }\href@noop {} {\bibfield  {journal} {\bibinfo  {journal} {Rev.
  Mod. Phys.}\ }\textbf {\bibinfo {volume} {81}},\ \bibinfo {pages} {163}
  (\bibinfo {year} {2009})}\BibitemShut {NoStop}%
\bibitem [{\citenamefont {Goulielmakis}\ \emph {et~al.}(2010)\citenamefont
  {Goulielmakis}, \citenamefont {Loh}, \citenamefont {Wirth}, \citenamefont
  {Santra}, \citenamefont {Rohringer}, \citenamefont {Yakovlev}, \citenamefont
  {Zherebtsov}, \citenamefont {Pfeifer}, \citenamefont {Azzeer}, \citenamefont
  {Kling} \emph {et~al.}}]{goulielmakis.2010.ATA.Kr}%
  \BibitemOpen
  \bibfield  {author} {\bibinfo {author} {\bibfnamefont {E.}~\bibnamefont
  {Goulielmakis}}, \bibinfo {author} {\bibfnamefont {Z.-H.}\ \bibnamefont
  {Loh}}, \bibinfo {author} {\bibfnamefont {A.}~\bibnamefont {Wirth}}, \bibinfo
  {author} {\bibfnamefont {R.}~\bibnamefont {Santra}}, \bibinfo {author}
  {\bibfnamefont {N.}~\bibnamefont {Rohringer}}, \bibinfo {author}
  {\bibfnamefont {V.~S.}\ \bibnamefont {Yakovlev}}, \bibinfo {author}
  {\bibfnamefont {S.}~\bibnamefont {Zherebtsov}}, \bibinfo {author}
  {\bibfnamefont {T.}~\bibnamefont {Pfeifer}}, \bibinfo {author} {\bibfnamefont
  {A.~M.}\ \bibnamefont {Azzeer}}, \bibinfo {author} {\bibfnamefont {M.~F.}\
  \bibnamefont {Kling}},  \emph {et~al.},\ }\href@noop {} {\bibfield  {journal}
  {\bibinfo  {journal} {Nature}\ }\textbf {\bibinfo {volume} {466}},\ \bibinfo
  {pages} {739} (\bibinfo {year} {2010})}\BibitemShut {NoStop}%
\bibitem [{\citenamefont {Wang}\ \emph {et~al.}(2010)\citenamefont {Wang},
  \citenamefont {Chini}, \citenamefont {Chen}, \citenamefont {Zhang},
  \citenamefont {He}, \citenamefont {Cheng}, \citenamefont {Wu}, \citenamefont
  {Thumm},\ and\ \citenamefont {Chang}}]{Wang.Chang.2010.ATA.Ar}%
  \BibitemOpen
  \bibfield  {author} {\bibinfo {author} {\bibfnamefont {H.}~\bibnamefont
  {Wang}}, \bibinfo {author} {\bibfnamefont {M.}~\bibnamefont {Chini}},
  \bibinfo {author} {\bibfnamefont {S.}~\bibnamefont {Chen}}, \bibinfo {author}
  {\bibfnamefont {C.-H.}\ \bibnamefont {Zhang}}, \bibinfo {author}
  {\bibfnamefont {F.}~\bibnamefont {He}}, \bibinfo {author} {\bibfnamefont
  {Y.}~\bibnamefont {Cheng}}, \bibinfo {author} {\bibfnamefont
  {Y.}~\bibnamefont {Wu}}, \bibinfo {author} {\bibfnamefont {U.}~\bibnamefont
  {Thumm}}, \ and\ \bibinfo {author} {\bibfnamefont {Z.}~\bibnamefont
  {Chang}},\ }\href@noop {} {\bibfield  {journal} {\bibinfo  {journal} {Phys.
  Rev. Lett.}\ }\textbf {\bibinfo {volume} {105}},\ \bibinfo {pages} {143002}
  (\bibinfo {year} {2010})}\BibitemShut {NoStop}%
\bibitem [{\citenamefont {Chini}\ \emph {et~al.}(2012)\citenamefont {Chini},
  \citenamefont {Zhao}, \citenamefont {Wang}, \citenamefont {Cheng},
  \citenamefont {Hu},\ and\ \citenamefont
  {Chang}}]{Chini.Chang.2012.ATA.ACstark}%
  \BibitemOpen
  \bibfield  {author} {\bibinfo {author} {\bibfnamefont {M.}~\bibnamefont
  {Chini}}, \bibinfo {author} {\bibfnamefont {B.}~\bibnamefont {Zhao}},
  \bibinfo {author} {\bibfnamefont {H.}~\bibnamefont {Wang}}, \bibinfo {author}
  {\bibfnamefont {Y.}~\bibnamefont {Cheng}}, \bibinfo {author} {\bibfnamefont
  {S.}~\bibnamefont {Hu}}, \ and\ \bibinfo {author} {\bibfnamefont
  {Z.}~\bibnamefont {Chang}},\ }\href@noop {} {\bibfield  {journal} {\bibinfo
  {journal} {Physical review letters}\ }\textbf {\bibinfo {volume} {109}},\
  \bibinfo {pages} {073601} (\bibinfo {year} {2012})}\BibitemShut {NoStop}%
\bibitem [{\citenamefont {Chen}\ \emph {et~al.}(2012)\citenamefont {Chen},
  \citenamefont {Bell}, \citenamefont {Beck}, \citenamefont {Mashiko},
  \citenamefont {Wu}, \citenamefont {Pfeiffer}, \citenamefont {Gaarde},
  \citenamefont {Neumark}, \citenamefont {Leone},\ and\ \citenamefont
  {Schafer}}]{chen.2012.ATA.LIS}%
  \BibitemOpen
  \bibfield  {author} {\bibinfo {author} {\bibfnamefont {S.}~\bibnamefont
  {Chen}}, \bibinfo {author} {\bibfnamefont {M.~J.}\ \bibnamefont {Bell}},
  \bibinfo {author} {\bibfnamefont {A.~R.}\ \bibnamefont {Beck}}, \bibinfo
  {author} {\bibfnamefont {H.}~\bibnamefont {Mashiko}}, \bibinfo {author}
  {\bibfnamefont {M.}~\bibnamefont {Wu}}, \bibinfo {author} {\bibfnamefont
  {A.~N.}\ \bibnamefont {Pfeiffer}}, \bibinfo {author} {\bibfnamefont {M.~B.}\
  \bibnamefont {Gaarde}}, \bibinfo {author} {\bibfnamefont {D.~M.}\
  \bibnamefont {Neumark}}, \bibinfo {author} {\bibfnamefont {S.~R.}\
  \bibnamefont {Leone}}, \ and\ \bibinfo {author} {\bibfnamefont {K.~J.}\
  \bibnamefont {Schafer}},\ }\href@noop {} {\bibfield  {journal} {\bibinfo
  {journal} {Phys. Rev. A}\ }\textbf {\bibinfo {volume} {86}},\ \bibinfo
  {pages} {063408} (\bibinfo {year} {2012})}\BibitemShut {NoStop}%
\bibitem [{\citenamefont {Holler}\ \emph {et~al.}(2011)\citenamefont {Holler},
  \citenamefont {Schapper}, \citenamefont {Gallmann},\ and\ \citenamefont
  {Keller}}]{Holler.Keller.2011.ATA.WPI}%
  \BibitemOpen
  \bibfield  {author} {\bibinfo {author} {\bibfnamefont {M.}~\bibnamefont
  {Holler}}, \bibinfo {author} {\bibfnamefont {F.}~\bibnamefont {Schapper}},
  \bibinfo {author} {\bibfnamefont {L.}~\bibnamefont {Gallmann}}, \ and\
  \bibinfo {author} {\bibfnamefont {U.}~\bibnamefont {Keller}},\ }\href@noop {}
  {\bibfield  {journal} {\bibinfo  {journal} {Phys. Rev. Lett.}\ }\textbf
  {\bibinfo {volume} {106}},\ \bibinfo {pages} {123601} (\bibinfo {year}
  {2011})}\BibitemShut {NoStop}%
\bibitem [{\citenamefont {Ott}\ \emph {et~al.}(2013)\citenamefont {Ott},
  \citenamefont {Kaldun}, \citenamefont {Raith}, \citenamefont {Meyer},
  \citenamefont {Laux}, \citenamefont {Evers}, \citenamefont {Keitel},
  \citenamefont {Greene},\ and\ \citenamefont
  {Pfeifer}}]{Pfeifer.2013.ATA.He.LorentzMeetsFano}%
  \BibitemOpen
  \bibfield  {author} {\bibinfo {author} {\bibfnamefont {C.}~\bibnamefont
  {Ott}}, \bibinfo {author} {\bibfnamefont {A.}~\bibnamefont {Kaldun}},
  \bibinfo {author} {\bibfnamefont {P.}~\bibnamefont {Raith}}, \bibinfo
  {author} {\bibfnamefont {K.}~\bibnamefont {Meyer}}, \bibinfo {author}
  {\bibfnamefont {M.}~\bibnamefont {Laux}}, \bibinfo {author} {\bibfnamefont
  {J.}~\bibnamefont {Evers}}, \bibinfo {author} {\bibfnamefont {C.~H.}\
  \bibnamefont {Keitel}}, \bibinfo {author} {\bibfnamefont {C.~H.}\
  \bibnamefont {Greene}}, \ and\ \bibinfo {author} {\bibfnamefont
  {T.}~\bibnamefont {Pfeifer}},\ }\href {\doibase 10.1126/science.1234407}
  {\bibfield  {journal} {\bibinfo  {journal} {Science}\ }\textbf {\bibinfo
  {volume} {340}},\ \bibinfo {pages} {716} (\bibinfo {year}
  {2013})}\BibitemShut {NoStop}%
\bibitem [{\citenamefont {Liao}\ \emph {et~al.}(2015)\citenamefont {Liao},
  \citenamefont {Sandhu}, \citenamefont {Camp}, \citenamefont {Schafer},\ and\
  \citenamefont {Gaarde}}]{LiaoPRL2015}%
  \BibitemOpen
  \bibfield  {author} {\bibinfo {author} {\bibfnamefont {C.-T.}\ \bibnamefont
  {Liao}}, \bibinfo {author} {\bibfnamefont {A.}~\bibnamefont {Sandhu}},
  \bibinfo {author} {\bibfnamefont {S.}~\bibnamefont {Camp}}, \bibinfo {author}
  {\bibfnamefont {K.~J.}\ \bibnamefont {Schafer}}, \ and\ \bibinfo {author}
  {\bibfnamefont {M.~B.}\ \bibnamefont {Gaarde}},\ }\href {\doibase
  10.1103/PhysRevLett.114.143002} {\bibfield  {journal} {\bibinfo  {journal}
  {Phys. Rev. Lett.}\ }\textbf {\bibinfo {volume} {114}},\ \bibinfo {pages}
  {143002} (\bibinfo {year} {2015})}\BibitemShut {NoStop}%
\bibitem [{\citenamefont {Liao}\ \emph {et~al.}(2016)\citenamefont {Liao},
  \citenamefont {Sandhu}, \citenamefont {Camp}, \citenamefont {Schafer},\ and\
  \citenamefont {Gaarde}}]{LiaoPRA2016}%
  \BibitemOpen
  \bibfield  {author} {\bibinfo {author} {\bibfnamefont {C.-T.}\ \bibnamefont
  {Liao}}, \bibinfo {author} {\bibfnamefont {A.}~\bibnamefont {Sandhu}},
  \bibinfo {author} {\bibfnamefont {S.}~\bibnamefont {Camp}}, \bibinfo {author}
  {\bibfnamefont {K.~J.}\ \bibnamefont {Schafer}}, \ and\ \bibinfo {author}
  {\bibfnamefont {M.~B.}\ \bibnamefont {Gaarde}},\ }\href@noop {} {\bibfield
  {journal} {\bibinfo  {journal} {Physical Review A}\ }\textbf {\bibinfo
  {volume} {93}},\ \bibinfo {pages} {033405} (\bibinfo {year}
  {2016})}\BibitemShut {NoStop}%
\bibitem [{\citenamefont {Warrick}\ \emph {et~al.}(2016)\citenamefont
  {Warrick}, \citenamefont {Cao}, \citenamefont {Neumark},\ and\ \citenamefont
  {Leone}}]{WarrickLeone.2016.ATA.N2}%
  \BibitemOpen
  \bibfield  {author} {\bibinfo {author} {\bibfnamefont {E.~R.}\ \bibnamefont
  {Warrick}}, \bibinfo {author} {\bibfnamefont {W.}~\bibnamefont {Cao}},
  \bibinfo {author} {\bibfnamefont {D.~M.}\ \bibnamefont {Neumark}}, \ and\
  \bibinfo {author} {\bibfnamefont {S.~R.}\ \bibnamefont {Leone}},\ }\href@noop
  {} {\bibfield  {journal} {\bibinfo  {journal} {The Journal of Physical
  Chemistry A}\ }\textbf {\bibinfo {volume} {120}},\ \bibinfo {pages} {3165}
  (\bibinfo {year} {2016})}\BibitemShut {NoStop}%
\bibitem [{\citenamefont {Reduzzi}\ \emph {et~al.}(2016)\citenamefont
  {Reduzzi}, \citenamefont {Chu}, \citenamefont {Feng}, \citenamefont
  {Dubrouil}, \citenamefont {Hummert}, \citenamefont {Calegari}, \citenamefont
  {Frassetto}, \citenamefont {Poletto}, \citenamefont {Kornilov}, \citenamefont
  {Nisoli} \emph {et~al.}}]{ReduzziSansone2016.ATA.N2}%
  \BibitemOpen
  \bibfield  {author} {\bibinfo {author} {\bibfnamefont {M.}~\bibnamefont
  {Reduzzi}}, \bibinfo {author} {\bibfnamefont {W.}~\bibnamefont {Chu}},
  \bibinfo {author} {\bibfnamefont {C.}~\bibnamefont {Feng}}, \bibinfo {author}
  {\bibfnamefont {A.}~\bibnamefont {Dubrouil}}, \bibinfo {author}
  {\bibfnamefont {J.}~\bibnamefont {Hummert}}, \bibinfo {author} {\bibfnamefont
  {F.}~\bibnamefont {Calegari}}, \bibinfo {author} {\bibfnamefont
  {F.}~\bibnamefont {Frassetto}}, \bibinfo {author} {\bibfnamefont
  {L.}~\bibnamefont {Poletto}}, \bibinfo {author} {\bibfnamefont
  {O.}~\bibnamefont {Kornilov}}, \bibinfo {author} {\bibfnamefont
  {M.}~\bibnamefont {Nisoli}},  \emph {et~al.},\ }\href@noop {} {\bibfield
  {journal} {\bibinfo  {journal} {Journal of Physics B: Atomic, Molecular and
  Optical Physics}\ }\textbf {\bibinfo {volume} {49}},\ \bibinfo {pages}
  {065102} (\bibinfo {year} {2016})}\BibitemShut {NoStop}%
\bibitem [{\citenamefont {Miroshnichenko}\ \emph {et~al.}(2010)\citenamefont
  {Miroshnichenko}, \citenamefont {Flach},\ and\ \citenamefont
  {Kivshar}}]{NanoStructure.Fano.2010}%
  \BibitemOpen
  \bibfield  {author} {\bibinfo {author} {\bibfnamefont {A.~E.}\ \bibnamefont
  {Miroshnichenko}}, \bibinfo {author} {\bibfnamefont {S.}~\bibnamefont
  {Flach}}, \ and\ \bibinfo {author} {\bibfnamefont {Y.~S.}\ \bibnamefont
  {Kivshar}},\ }\href {\doibase 10.1103/RevModPhys.82.2257} {\bibfield
  {journal} {\bibinfo  {journal} {Rev. Mod. Phys.}\ }\textbf {\bibinfo {volume}
  {82}},\ \bibinfo {pages} {2257} (\bibinfo {year} {2010})}\BibitemShut
  {NoStop}%
\bibitem [{\citenamefont {Fano}(1961)}]{Fano_1961}%
  \BibitemOpen
  \bibfield  {author} {\bibinfo {author} {\bibfnamefont {U.}~\bibnamefont
  {Fano}},\ }\href@noop {} {\bibfield  {journal} {\bibinfo  {journal} {Phys.
  Rev.}\ }\textbf {\bibinfo {volume} {124}},\ \bibinfo {pages} {1866} (\bibinfo
  {year} {1961})}\BibitemShut {NoStop}%
\bibitem [{\citenamefont {Holland}\ \emph {et~al.}(1993)\citenamefont
  {Holland}, \citenamefont {Shaw}, \citenamefont {McSweeney}, \citenamefont
  {MacDonald}, \citenamefont {Hopkirk},\ and\ \citenamefont {Hayes}}]{Holland}%
  \BibitemOpen
  \bibfield  {author} {\bibinfo {author} {\bibfnamefont {D.}~\bibnamefont
  {Holland}}, \bibinfo {author} {\bibfnamefont {D.}~\bibnamefont {Shaw}},
  \bibinfo {author} {\bibfnamefont {S.}~\bibnamefont {McSweeney}}, \bibinfo
  {author} {\bibfnamefont {M.}~\bibnamefont {MacDonald}}, \bibinfo {author}
  {\bibfnamefont {A.}~\bibnamefont {Hopkirk}}, \ and\ \bibinfo {author}
  {\bibfnamefont {M.}~\bibnamefont {Hayes}},\ }\href@noop {} {\bibfield
  {journal} {\bibinfo  {journal} {Chemical physics}\ }\textbf {\bibinfo
  {volume} {173}},\ \bibinfo {pages} {315} (\bibinfo {year}
  {1993})}\BibitemShut {NoStop}%
\bibitem [{\citenamefont {Demekhin}\ \emph {et~al.}(2007)\citenamefont
  {Demekhin}, \citenamefont {Omelyanenko}, \citenamefont {Lagutin},
  \citenamefont {Sukhorukov}, \citenamefont {Werner}, \citenamefont
  {Ehresmann}, \citenamefont {Schartner},\ and\ \citenamefont
  {Schmoranzer}}]{Demekhin}%
  \BibitemOpen
  \bibfield  {author} {\bibinfo {author} {\bibfnamefont {P.~V.}\ \bibnamefont
  {Demekhin}}, \bibinfo {author} {\bibfnamefont {D.}~\bibnamefont
  {Omelyanenko}}, \bibinfo {author} {\bibfnamefont {B.}~\bibnamefont
  {Lagutin}}, \bibinfo {author} {\bibfnamefont {V.}~\bibnamefont {Sukhorukov}},
  \bibinfo {author} {\bibfnamefont {L.}~\bibnamefont {Werner}}, \bibinfo
  {author} {\bibfnamefont {A.}~\bibnamefont {Ehresmann}}, \bibinfo {author}
  {\bibfnamefont {K.-H.}\ \bibnamefont {Schartner}}, \ and\ \bibinfo {author}
  {\bibfnamefont {H.}~\bibnamefont {Schmoranzer}},\ }\href@noop {} {\bibfield
  {journal} {\bibinfo  {journal} {Optics and spectroscopy}\ }\textbf {\bibinfo
  {volume} {102}},\ \bibinfo {pages} {318} (\bibinfo {year}
  {2007})}\BibitemShut {NoStop}%
\bibitem [{\citenamefont {Gilmore}(1965)}]{Gilmore1965}%
  \BibitemOpen
  \bibfield  {author} {\bibinfo {author} {\bibfnamefont {F.~R.}\ \bibnamefont
  {Gilmore}},\ }\href {\doibase http://dx.doi.org/10.1016/0022-4073(65)90072-5}
  {\bibfield  {journal} {\bibinfo  {journal} {Journal of Quantitative
  Spectroscopy and Radiative Transfer}\ }\textbf {\bibinfo {volume} {5}},\
  \bibinfo {pages} {369 } (\bibinfo {year} {1965})}\BibitemShut {NoStop}%
\bibitem [{\citenamefont {Baltzer}\ \emph
  {et~al.}(1992{\natexlab{a}})\citenamefont {Baltzer}, \citenamefont
  {Wannberg}, \citenamefont {Karlsson}, \citenamefont {Carlsson~G\"othe},\ and\
  \citenamefont {Larsson}}]{Larsson1992}%
  \BibitemOpen
  \bibfield  {author} {\bibinfo {author} {\bibfnamefont {P.}~\bibnamefont
  {Baltzer}}, \bibinfo {author} {\bibfnamefont {B.}~\bibnamefont {Wannberg}},
  \bibinfo {author} {\bibfnamefont {L.}~\bibnamefont {Karlsson}}, \bibinfo
  {author} {\bibfnamefont {M.}~\bibnamefont {Carlsson~G\"othe}}, \ and\
  \bibinfo {author} {\bibfnamefont {M.}~\bibnamefont {Larsson}},\ }\href
  {\doibase 10.1103/PhysRevA.45.4374} {\bibfield  {journal} {\bibinfo
  {journal} {Phys. Rev. A}\ }\textbf {\bibinfo {volume} {45}},\ \bibinfo
  {pages} {4374} (\bibinfo {year} {1992}{\natexlab{a}})}\BibitemShut {NoStop}%
\bibitem [{\citenamefont {Ehresmann}\ \emph {et~al.}(2004)\citenamefont
  {Ehresmann}, \citenamefont {Werner}, \citenamefont {Klumpp}, \citenamefont
  {Schmoranzer}, \citenamefont {Demekhin}, \citenamefont {Lagutin},
  \citenamefont {Sukhorukov}, \citenamefont {Mickat}, \citenamefont {Kammer},
  \citenamefont {Zimmermann} \emph {et~al.}}]{Ehresmann2004.O2}%
  \BibitemOpen
  \bibfield  {author} {\bibinfo {author} {\bibfnamefont {A.}~\bibnamefont
  {Ehresmann}}, \bibinfo {author} {\bibfnamefont {L.}~\bibnamefont {Werner}},
  \bibinfo {author} {\bibfnamefont {S.}~\bibnamefont {Klumpp}}, \bibinfo
  {author} {\bibfnamefont {H.}~\bibnamefont {Schmoranzer}}, \bibinfo {author}
  {\bibfnamefont {P.~V.}\ \bibnamefont {Demekhin}}, \bibinfo {author}
  {\bibfnamefont {B.}~\bibnamefont {Lagutin}}, \bibinfo {author} {\bibfnamefont
  {V.}~\bibnamefont {Sukhorukov}}, \bibinfo {author} {\bibfnamefont
  {S.}~\bibnamefont {Mickat}}, \bibinfo {author} {\bibfnamefont
  {S.}~\bibnamefont {Kammer}}, \bibinfo {author} {\bibfnamefont
  {B.}~\bibnamefont {Zimmermann}},  \emph {et~al.},\ }\href@noop {} {\bibfield
  {journal} {\bibinfo  {journal} {J. Phys. B: At. Mol. Opt. Phys.}\ }\textbf
  {\bibinfo {volume} {37}},\ \bibinfo {pages} {4405} (\bibinfo {year}
  {2004})}\BibitemShut {NoStop}%
\bibitem [{\citenamefont {Ott}\ \emph {et~al.}(2014)\citenamefont {Ott},
  \citenamefont {Kaldun}, \citenamefont {Argenti}, \citenamefont {Raith},
  \citenamefont {Meyer}, \citenamefont {Laux}, \citenamefont {Zhang},
  \citenamefont {Bl{\"a}ttermann}, \citenamefont {Hagstotz}, \citenamefont
  {Ding} \emph {et~al.}}]{ott.2014.ATA.Recon}%
  \BibitemOpen
  \bibfield  {author} {\bibinfo {author} {\bibfnamefont {C.}~\bibnamefont
  {Ott}}, \bibinfo {author} {\bibfnamefont {A.}~\bibnamefont {Kaldun}},
  \bibinfo {author} {\bibfnamefont {L.}~\bibnamefont {Argenti}}, \bibinfo
  {author} {\bibfnamefont {P.}~\bibnamefont {Raith}}, \bibinfo {author}
  {\bibfnamefont {K.}~\bibnamefont {Meyer}}, \bibinfo {author} {\bibfnamefont
  {M.}~\bibnamefont {Laux}}, \bibinfo {author} {\bibfnamefont {Y.}~\bibnamefont
  {Zhang}}, \bibinfo {author} {\bibfnamefont {A.}~\bibnamefont
  {Bl{\"a}ttermann}}, \bibinfo {author} {\bibfnamefont {S.}~\bibnamefont
  {Hagstotz}}, \bibinfo {author} {\bibfnamefont {T.}~\bibnamefont {Ding}},
  \emph {et~al.},\ }\href@noop {} {\bibfield  {journal} {\bibinfo  {journal}
  {Nature}\ }\textbf {\bibinfo {volume} {516}},\ \bibinfo {pages} {374}
  (\bibinfo {year} {2014})}\BibitemShut {NoStop}%
\bibitem [{\citenamefont {Wu}\ \emph {et~al.}(2016)\citenamefont {Wu},
  \citenamefont {Chen}, \citenamefont {Camp}, \citenamefont {Schafer},\ and\
  \citenamefont {Gaarde}}]{wu.2016.ATA.review}%
  \BibitemOpen
  \bibfield  {author} {\bibinfo {author} {\bibfnamefont {M.}~\bibnamefont
  {Wu}}, \bibinfo {author} {\bibfnamefont {S.}~\bibnamefont {Chen}}, \bibinfo
  {author} {\bibfnamefont {S.}~\bibnamefont {Camp}}, \bibinfo {author}
  {\bibfnamefont {K.~J.}\ \bibnamefont {Schafer}}, \ and\ \bibinfo {author}
  {\bibfnamefont {M.~B.}\ \bibnamefont {Gaarde}},\ }\href@noop {} {\bibfield
  {journal} {\bibinfo  {journal} {J. Phys. B: Atom. Mol. Phys.}\ }\textbf
  {\bibinfo {volume} {49}},\ \bibinfo {pages} {062003} (\bibinfo {year}
  {2016})}\BibitemShut {NoStop}%
\bibitem [{\citenamefont {Stratmann}\ and\ \citenamefont
  {Lucchese}(1995)}]{Stratmann1995}%
  \BibitemOpen
  \bibfield  {author} {\bibinfo {author} {\bibfnamefont {R.~E.}\ \bibnamefont
  {Stratmann}}\ and\ \bibinfo {author} {\bibfnamefont {R.~R.}\ \bibnamefont
  {Lucchese}},\ }\href@noop {} {\bibfield  {journal} {\bibinfo  {journal}
  {Journal of Chemical Physics}\ }\textbf {\bibinfo {volume} {102}},\ \bibinfo
  {pages} {8493} (\bibinfo {year} {1995})}\BibitemShut {NoStop}%
\bibitem [{\citenamefont {Stratmann}\ \emph {et~al.}(1996)\citenamefont
  {Stratmann}, \citenamefont {Zurales},\ and\ \citenamefont
  {Lucchese}}]{Stratmann1996}%
  \BibitemOpen
  \bibfield  {author} {\bibinfo {author} {\bibfnamefont {R.~E.}\ \bibnamefont
  {Stratmann}}, \bibinfo {author} {\bibfnamefont {R.~W.}\ \bibnamefont
  {Zurales}}, \ and\ \bibinfo {author} {\bibfnamefont {R.~R.}\ \bibnamefont
  {Lucchese}},\ }\href@noop {} {\bibfield  {journal} {\bibinfo  {journal}
  {Journal of Chemical Physics}\ }\textbf {\bibinfo {volume} {104}},\ \bibinfo
  {pages} {8989} (\bibinfo {year} {1996})}\BibitemShut {NoStop}%
\bibitem [{\citenamefont {Dunning}(1989)}]{Dunning1989}%
  \BibitemOpen
  \bibfield  {author} {\bibinfo {author} {\bibfnamefont {J.}~\bibnamefont
  {Dunning}, \bibfnamefont {Thom~H.}},\ }\href@noop {} {\bibfield  {journal}
  {\bibinfo  {journal} {Journal of Chemical Physics}\ }\textbf {\bibinfo
  {volume} {90}},\ \bibinfo {pages} {1007} (\bibinfo {year}
  {1989})}\BibitemShut {NoStop}%
\bibitem [{\citenamefont {Kendall}\ \emph {et~al.}(1992)\citenamefont
  {Kendall}, \citenamefont {Dunning},\ and\ \citenamefont
  {Harrison}}]{Kendall1992}%
  \BibitemOpen
  \bibfield  {author} {\bibinfo {author} {\bibfnamefont {R.~A.}\ \bibnamefont
  {Kendall}}, \bibinfo {author} {\bibfnamefont {J.}~\bibnamefont {Dunning},
  \bibfnamefont {Thom~H.}}, \ and\ \bibinfo {author} {\bibfnamefont {R.~J.}\
  \bibnamefont {Harrison}},\ }\href@noop {} {\bibfield  {journal} {\bibinfo
  {journal} {Journal of Chemical Physics}\ }\textbf {\bibinfo {volume} {96}},\
  \bibinfo {pages} {6796} (\bibinfo {year} {1992})}\BibitemShut {NoStop}%
\bibitem [{\citenamefont {Baltzer}\ \emph
  {et~al.}(1992{\natexlab{b}})\citenamefont {Baltzer}, \citenamefont
  {Wannberg}, \citenamefont {Karlsson}, \citenamefont {Gothe},\ and\
  \citenamefont {Larsson}}]{Baltzer1992}%
  \BibitemOpen
  \bibfield  {author} {\bibinfo {author} {\bibfnamefont {P.}~\bibnamefont
  {Baltzer}}, \bibinfo {author} {\bibfnamefont {B.}~\bibnamefont {Wannberg}},
  \bibinfo {author} {\bibfnamefont {L.}~\bibnamefont {Karlsson}}, \bibinfo
  {author} {\bibfnamefont {M.~C.}\ \bibnamefont {Gothe}}, \ and\ \bibinfo
  {author} {\bibfnamefont {M.}~\bibnamefont {Larsson}},\ }\href@noop {}
  {\bibfield  {journal} {\bibinfo  {journal} {Physical Review A}\ }\textbf
  {\bibinfo {volume} {45}},\ \bibinfo {pages} {4374} (\bibinfo {year}
  {1992}{\natexlab{b}})}\BibitemShut {NoStop}%
\bibitem [{\citenamefont {Wu}(1987)}]{Wu1987}%
  \BibitemOpen
  \bibfield  {author} {\bibinfo {author} {\bibfnamefont {C.~R.}\ \bibnamefont
  {Wu}},\ }\href {\doibase http://dx.doi.org/10.1016/0022-4073(87)90115-4}
  {\bibfield  {journal} {\bibinfo  {journal} {Journal of Quantitative
  Spectroscopy and Radiative Transfer}\ }\textbf {\bibinfo {volume} {37}},\
  \bibinfo {pages} {1 } (\bibinfo {year} {1987})}\BibitemShut {NoStop}%
\bibitem [{\citenamefont {Alon}\ \emph {et~al.}(2007)\citenamefont {Alon},
  \citenamefont {Streitsov},\ and\ \citenamefont {Cederbaum}}]{Cederbaum2007}%
  \BibitemOpen
  \bibfield  {author} {\bibinfo {author} {\bibfnamefont {O.~E.}\ \bibnamefont
  {Alon}}, \bibinfo {author} {\bibfnamefont {A.~I.}\ \bibnamefont {Streitsov}},
  \ and\ \bibinfo {author} {\bibfnamefont {L.~S.}\ \bibnamefont {Cederbaum}},\
  }\href@noop {} {\bibfield  {journal} {\bibinfo  {journal} {J. Chem. Phys.}\
  }\textbf {\bibinfo {volume} {127}},\ \bibinfo {pages} {154103} (\bibinfo
  {year} {2007})}\BibitemShut {NoStop}%
\bibitem [{\citenamefont {Caillat}\ \emph {et~al.}(2005)\citenamefont {Caillat}
  \emph {et~al.}}]{Scrinzi_MCTDHF_2005}%
  \BibitemOpen
  \bibfield  {author} {\bibinfo {author} {\bibfnamefont {J.}~\bibnamefont
  {Caillat}} \emph {et~al.},\ }\href {\doibase 10.1103/PhysRevA.71.012712}
  {\bibfield  {journal} {\bibinfo  {journal} {Phys. Rev. A}\ }\textbf {\bibinfo
  {volume} {71}},\ \bibinfo {pages} {012712} (\bibinfo {year}
  {2005})}\BibitemShut {NoStop}%
\bibitem [{\citenamefont {Kato}\ and\ \citenamefont
  {Kono}(2009)}]{Kato_Kono2009}%
  \BibitemOpen
  \bibfield  {author} {\bibinfo {author} {\bibfnamefont {T.}~\bibnamefont
  {Kato}}\ and\ \bibinfo {author} {\bibfnamefont {H.}~\bibnamefont {Kono}},\
  }\href@noop {} {\bibfield  {journal} {\bibinfo  {journal} {Chem. Phys.}\
  }\textbf {\bibinfo {volume} {366}},\ \bibinfo {pages} {46} (\bibinfo {year}
  {2009})}\BibitemShut {NoStop}%
\bibitem [{\citenamefont {Ulusoy}\ and\ \citenamefont
  {Nest}(2012)}]{nest_lih2012}%
  \BibitemOpen
  \bibfield  {author} {\bibinfo {author} {\bibfnamefont {I.~S.}\ \bibnamefont
  {Ulusoy}}\ and\ \bibinfo {author} {\bibfnamefont {M.}~\bibnamefont {Nest}},\
  }\href@noop {} {\bibfield  {journal} {\bibinfo  {journal} {J. Chem. Phys.}\
  }\textbf {\bibinfo {volume} {136}},\ \bibinfo {pages} {054112} (\bibinfo
  {year} {2012})}\BibitemShut {NoStop}%
\bibitem [{\citenamefont {Miranda}\ \emph {et~al.}(2011)\citenamefont
  {Miranda}, \citenamefont {Fisher}, \citenamefont {Stella},\ and\
  \citenamefont {Horsfield}}]{Miranda_2011}%
  \BibitemOpen
  \bibfield  {author} {\bibinfo {author} {\bibfnamefont {R.~P.}\ \bibnamefont
  {Miranda}}, \bibinfo {author} {\bibfnamefont {A.~J.}\ \bibnamefont {Fisher}},
  \bibinfo {author} {\bibfnamefont {L.}~\bibnamefont {Stella}}, \ and\ \bibinfo
  {author} {\bibfnamefont {A.~P.}\ \bibnamefont {Horsfield}},\ }\href@noop {}
  {\bibfield  {journal} {\bibinfo  {journal} {J. Chem. Phys.}\ }\textbf
  {\bibinfo {volume} {134}},\ \bibinfo {pages} {244101} (\bibinfo {year}
  {2011})}\BibitemShut {NoStop}%
\bibitem [{\citenamefont {Miyagi}\ and\ \citenamefont
  {Madsen}(2013)}]{Madsen_2013}%
  \BibitemOpen
  \bibfield  {author} {\bibinfo {author} {\bibfnamefont {H.}~\bibnamefont
  {Miyagi}}\ and\ \bibinfo {author} {\bibfnamefont {L.~B.}\ \bibnamefont
  {Madsen}},\ }\href {\doibase 10.1103/PhysRevA.87.062511} {\bibfield
  {journal} {\bibinfo  {journal} {Phys. Rev. A}\ }\textbf {\bibinfo {volume}
  {87}},\ \bibinfo {pages} {062511} (\bibinfo {year} {2013})}\BibitemShut
  {NoStop}%
\bibitem [{\citenamefont {Sato}\ and\ \citenamefont
  {Ishikawa}(2013)}]{Sato_2013}%
  \BibitemOpen
  \bibfield  {author} {\bibinfo {author} {\bibfnamefont {T.}~\bibnamefont
  {Sato}}\ and\ \bibinfo {author} {\bibfnamefont {K.~L.}\ \bibnamefont
  {Ishikawa}},\ }\href {\doibase 10.1103/PhysRevA.88.023402} {\bibfield
  {journal} {\bibinfo  {journal} {Phys. Rev. A}\ }\textbf {\bibinfo {volume}
  {88}},\ \bibinfo {pages} {023402} (\bibinfo {year} {2013})}\BibitemShut
  {NoStop}%
\bibitem [{\citenamefont {Tannor}(2007)}]{TannorBook}%
  \BibitemOpen
  \bibfield  {author} {\bibinfo {author} {\bibfnamefont {D.~J.}\ \bibnamefont
  {Tannor}},\ }\href@noop {} {\emph {\bibinfo {title} {Introduction to Quantum
  Mechanics: A Time Dependent Perspective}}}\ (\bibinfo  {publisher}
  {University Science Press},\ \bibinfo {address} {Sausalito},\ \bibinfo {year}
  {2007})\BibitemShut {NoStop}%
\bibitem [{\citenamefont {Gaarde}\ \emph {et~al.}(2011)\citenamefont {Gaarde},
  \citenamefont {Buth}, \citenamefont {Tate},\ and\ \citenamefont
  {Schafer}}]{Gaarde2011.ATA.Response}%
  \BibitemOpen
  \bibfield  {author} {\bibinfo {author} {\bibfnamefont {M.~B.}\ \bibnamefont
  {Gaarde}}, \bibinfo {author} {\bibfnamefont {C.}~\bibnamefont {Buth}},
  \bibinfo {author} {\bibfnamefont {J.~T.}\ \bibnamefont {Tate}}, \ and\
  \bibinfo {author} {\bibfnamefont {K.~J.}\ \bibnamefont {Schafer}},\
  }\href@noop {} {\bibfield  {journal} {\bibinfo  {journal} {Phys. Rev. A}\
  }\textbf {\bibinfo {volume} {83}},\ \bibinfo {pages} {013419} (\bibinfo
  {year} {2011})}\BibitemShut {NoStop}%
\bibitem [{\citenamefont {Chu}\ and\ \citenamefont {Lin}(2013)}]{CDLin2013}%
  \BibitemOpen
  \bibfield  {author} {\bibinfo {author} {\bibfnamefont {W.-C.}\ \bibnamefont
  {Chu}}\ and\ \bibinfo {author} {\bibfnamefont {C.}~\bibnamefont {Lin}},\
  }\href@noop {} {\bibfield  {journal} {\bibinfo  {journal} {Physical Review
  A}\ }\textbf {\bibinfo {volume} {87}},\ \bibinfo {pages} {013415} (\bibinfo
  {year} {2013})}\BibitemShut {NoStop}%
\bibitem [{\citenamefont {Cubric}\ \emph {et~al.}(1993)\citenamefont {Cubric},
  \citenamefont {Wills}, \citenamefont {Comer},\ and\ \citenamefont
  {Ukai}}]{cubric1993.O2}%
  \BibitemOpen
  \bibfield  {author} {\bibinfo {author} {\bibfnamefont {D.}~\bibnamefont
  {Cubric}}, \bibinfo {author} {\bibfnamefont {A.}~\bibnamefont {Wills}},
  \bibinfo {author} {\bibfnamefont {J.}~\bibnamefont {Comer}}, \ and\ \bibinfo
  {author} {\bibfnamefont {M.}~\bibnamefont {Ukai}},\ }\href@noop {} {\bibfield
   {journal} {\bibinfo  {journal} {J. Phys. B: At. Mol. Opt. Phys.}\ }\textbf
  {\bibinfo {volume} {26}},\ \bibinfo {pages} {3081} (\bibinfo {year}
  {1993})}\BibitemShut {NoStop}%
\bibitem [{\citenamefont {Liebel}\ \emph {et~al.}(2000)\citenamefont {Liebel},
  \citenamefont {Lauer}, \citenamefont {Vollweiler}, \citenamefont
  {M{\"u}ller-Albrecht}, \citenamefont {Ehresmann}, \citenamefont
  {Schmoranzer}, \citenamefont {Mentzel}, \citenamefont {Schartner},\ and\
  \citenamefont {Wilhelmi}}]{liebel2000}%
  \BibitemOpen
  \bibfield  {author} {\bibinfo {author} {\bibfnamefont {H.}~\bibnamefont
  {Liebel}}, \bibinfo {author} {\bibfnamefont {S.}~\bibnamefont {Lauer}},
  \bibinfo {author} {\bibfnamefont {F.}~\bibnamefont {Vollweiler}}, \bibinfo
  {author} {\bibfnamefont {R.}~\bibnamefont {M{\"u}ller-Albrecht}}, \bibinfo
  {author} {\bibfnamefont {A.}~\bibnamefont {Ehresmann}}, \bibinfo {author}
  {\bibfnamefont {H.}~\bibnamefont {Schmoranzer}}, \bibinfo {author}
  {\bibfnamefont {G.}~\bibnamefont {Mentzel}}, \bibinfo {author} {\bibfnamefont
  {K.-H.}\ \bibnamefont {Schartner}}, \ and\ \bibinfo {author} {\bibfnamefont
  {O.}~\bibnamefont {Wilhelmi}},\ }\href@noop {} {\bibfield  {journal}
  {\bibinfo  {journal} {Physics Letters A}\ }\textbf {\bibinfo {volume}
  {267}},\ \bibinfo {pages} {357} (\bibinfo {year} {2000})}\BibitemShut
  {NoStop}%
\bibitem [{\citenamefont {Liebel}\ \emph {et~al.}(2002)\citenamefont {Liebel},
  \citenamefont {Ehresmann}, \citenamefont {Schmoranzer}, \citenamefont
  {Demekhin}, \citenamefont {Lagutin},\ and\ \citenamefont
  {Sukhorukov}}]{liebel2002}%
  \BibitemOpen
  \bibfield  {author} {\bibinfo {author} {\bibfnamefont {H.}~\bibnamefont
  {Liebel}}, \bibinfo {author} {\bibfnamefont {A.}~\bibnamefont {Ehresmann}},
  \bibinfo {author} {\bibfnamefont {H.}~\bibnamefont {Schmoranzer}}, \bibinfo
  {author} {\bibfnamefont {P.~V.}\ \bibnamefont {Demekhin}}, \bibinfo {author}
  {\bibfnamefont {B.}~\bibnamefont {Lagutin}}, \ and\ \bibinfo {author}
  {\bibfnamefont {V.}~\bibnamefont {Sukhorukov}},\ }\href@noop {} {\bibfield
  {journal} {\bibinfo  {journal} {Journal of Physics B: Atomic, Molecular and
  Optical Physics}\ }\textbf {\bibinfo {volume} {35}},\ \bibinfo {pages} {895}
  (\bibinfo {year} {2002})}\BibitemShut {NoStop}%
\bibitem [{\citenamefont {Hikosaka}\ \emph {et~al.}(2003)\citenamefont
  {Hikosaka}, \citenamefont {Lablanquie}, \citenamefont {Ahmad}, \citenamefont
  {Hall}, \citenamefont {Lambourne}, \citenamefont {Penent},\ and\
  \citenamefont {Eland}}]{hikosaka2003}%
  \BibitemOpen
  \bibfield  {author} {\bibinfo {author} {\bibfnamefont {Y.}~\bibnamefont
  {Hikosaka}}, \bibinfo {author} {\bibfnamefont {P.}~\bibnamefont
  {Lablanquie}}, \bibinfo {author} {\bibfnamefont {M.}~\bibnamefont {Ahmad}},
  \bibinfo {author} {\bibfnamefont {R.}~\bibnamefont {Hall}}, \bibinfo {author}
  {\bibfnamefont {J.}~\bibnamefont {Lambourne}}, \bibinfo {author}
  {\bibfnamefont {F.}~\bibnamefont {Penent}}, \ and\ \bibinfo {author}
  {\bibfnamefont {J.}~\bibnamefont {Eland}},\ }\href@noop {} {\bibfield
  {journal} {\bibinfo  {journal} {J. Phys. B: At. Mol. Opt. Phys.}\ }\textbf
  {\bibinfo {volume} {36}},\ \bibinfo {pages} {4311} (\bibinfo {year}
  {2003})}\BibitemShut {NoStop}%
\bibitem [{\citenamefont {Doughty}\ \emph {et~al.}(2012)\citenamefont
  {Doughty}, \citenamefont {Koh}, \citenamefont {Haber},\ and\ \citenamefont
  {Leone}}]{doughty.Leone.2012.O2.VMI}%
  \BibitemOpen
  \bibfield  {author} {\bibinfo {author} {\bibfnamefont {B.}~\bibnamefont
  {Doughty}}, \bibinfo {author} {\bibfnamefont {C.~J.}\ \bibnamefont {Koh}},
  \bibinfo {author} {\bibfnamefont {L.~H.}\ \bibnamefont {Haber}}, \ and\
  \bibinfo {author} {\bibfnamefont {S.~R.}\ \bibnamefont {Leone}},\ }\href@noop
  {} {\bibfield  {journal} {\bibinfo  {journal} {J. Chem. Phys.}\ }\textbf
  {\bibinfo {volume} {136}},\ \bibinfo {pages} {214303} (\bibinfo {year}
  {2012})}\BibitemShut {NoStop}%
\bibitem [{\citenamefont {Lefebvre-Brion}\ and\ \citenamefont
  {Field}(2004)}]{Lefebvre-Brion2004}%
  \BibitemOpen
  \bibfield  {author} {\bibinfo {author} {\bibfnamefont {H.}~\bibnamefont
  {Lefebvre-Brion}}\ and\ \bibinfo {author} {\bibfnamefont {R.~W.}\
  \bibnamefont {Field}},\ }\href@noop {} {\emph {\bibinfo {title} {The Spectra
  and Dynamics of Diatomic Molecules}}}\ (\bibinfo  {publisher} {Elsevier
  Inc.},\ \bibinfo {year} {2004})\ p.\ \bibinfo {pages} {572}\BibitemShut
  {NoStop}%
\bibitem [{\citenamefont {Li}\ \emph {et~al.}(2015{\natexlab{a}})\citenamefont
  {Li}, \citenamefont {Bernhardt}, \citenamefont {Beck}, \citenamefont
  {Warrick}, \citenamefont {Pfeiffer}, \citenamefont {Bell}, \citenamefont
  {Haxton}, \citenamefont {McCurdy}, \citenamefont {Neumark},\ and\
  \citenamefont {Leone}}]{Li2015}%
  \BibitemOpen
  \bibfield  {author} {\bibinfo {author} {\bibfnamefont {X.}~\bibnamefont
  {Li}}, \bibinfo {author} {\bibfnamefont {B.}~\bibnamefont {Bernhardt}},
  \bibinfo {author} {\bibfnamefont {A.~R.}\ \bibnamefont {Beck}}, \bibinfo
  {author} {\bibfnamefont {E.~R.}\ \bibnamefont {Warrick}}, \bibinfo {author}
  {\bibfnamefont {A.~N.}\ \bibnamefont {Pfeiffer}}, \bibinfo {author}
  {\bibfnamefont {M.~J.}\ \bibnamefont {Bell}}, \bibinfo {author}
  {\bibfnamefont {D.~J.}\ \bibnamefont {Haxton}}, \bibinfo {author}
  {\bibfnamefont {C.~W.}\ \bibnamefont {McCurdy}}, \bibinfo {author}
  {\bibfnamefont {D.~M.}\ \bibnamefont {Neumark}}, \ and\ \bibinfo {author}
  {\bibfnamefont {S.~R.}\ \bibnamefont {Leone}},\ }\href@noop {} {\bibfield
  {journal} {\bibinfo  {journal} {Journal of Physics B: Atomic, Molecular and
  Optical Physics}\ }\textbf {\bibinfo {volume} {48}},\ \bibinfo {pages}
  {125601} (\bibinfo {year} {2015}{\natexlab{a}})}\BibitemShut {NoStop}%
\bibitem [{\citenamefont {Padmanabhan}\ \emph {et~al.}(2010)\citenamefont
  {Padmanabhan}, \citenamefont {MacDonald}, \citenamefont {Ryan}, \citenamefont
  {Zuin},\ and\ \citenamefont {Reddish}}]{Padmanabhan2010.Lifetime}%
  \BibitemOpen
  \bibfield  {author} {\bibinfo {author} {\bibfnamefont {A.}~\bibnamefont
  {Padmanabhan}}, \bibinfo {author} {\bibfnamefont {M.}~\bibnamefont
  {MacDonald}}, \bibinfo {author} {\bibfnamefont {C.}~\bibnamefont {Ryan}},
  \bibinfo {author} {\bibfnamefont {L.}~\bibnamefont {Zuin}}, \ and\ \bibinfo
  {author} {\bibfnamefont {T.}~\bibnamefont {Reddish}},\ }\href@noop {}
  {\bibfield  {journal} {\bibinfo  {journal} {Journal of Physics B: Atomic,
  Molecular and Optical Physics}\ }\textbf {\bibinfo {volume} {43}},\ \bibinfo
  {pages} {165204} (\bibinfo {year} {2010})}\BibitemShut {NoStop}%
\bibitem [{\citenamefont {B\ae{}kh\o{}j}\ \emph {et~al.}(2015)\citenamefont
  {B\ae{}kh\o{}j}, \citenamefont {Yue},\ and\ \citenamefont
  {Madsen}}]{Madsen2015_ATA_H2}%
  \BibitemOpen
  \bibfield  {author} {\bibinfo {author} {\bibfnamefont {J.~E.}\ \bibnamefont
  {B\ae{}kh\o{}j}}, \bibinfo {author} {\bibfnamefont {L.}~\bibnamefont {Yue}},
  \ and\ \bibinfo {author} {\bibfnamefont {L.~B.}\ \bibnamefont {Madsen}},\
  }\href {\doibase 10.1103/PhysRevA.91.043408} {\bibfield  {journal} {\bibinfo
  {journal} {Phys. Rev. A}\ }\textbf {\bibinfo {volume} {91}},\ \bibinfo
  {pages} {043408} (\bibinfo {year} {2015})}\BibitemShut {NoStop}%
\bibitem [{\citenamefont {Bernhardt}\ \emph {et~al.}(2014)\citenamefont
  {Bernhardt}, \citenamefont {Beck}, \citenamefont {Li}, \citenamefont
  {Warrick}, \citenamefont {Bell}, \citenamefont {Haxton}, \citenamefont
  {McCurdy}, \citenamefont {Neumark},\ and\ \citenamefont
  {Leone}}]{Bernhardt_Xenon_2014PRA}%
  \BibitemOpen
  \bibfield  {author} {\bibinfo {author} {\bibfnamefont {B.}~\bibnamefont
  {Bernhardt}}, \bibinfo {author} {\bibfnamefont {A.~R.}\ \bibnamefont {Beck}},
  \bibinfo {author} {\bibfnamefont {X.}~\bibnamefont {Li}}, \bibinfo {author}
  {\bibfnamefont {E.~R.}\ \bibnamefont {Warrick}}, \bibinfo {author}
  {\bibfnamefont {M.~J.}\ \bibnamefont {Bell}}, \bibinfo {author}
  {\bibfnamefont {D.~J.}\ \bibnamefont {Haxton}}, \bibinfo {author}
  {\bibfnamefont {C.~W.}\ \bibnamefont {McCurdy}}, \bibinfo {author}
  {\bibfnamefont {D.~M.}\ \bibnamefont {Neumark}}, \ and\ \bibinfo {author}
  {\bibfnamefont {S.~R.}\ \bibnamefont {Leone}},\ }\href {\doibase
  10.1103/PhysRevA.89.023408} {\bibfield  {journal} {\bibinfo  {journal} {Phys.
  Rev. A}\ }\textbf {\bibinfo {volume} {89}},\ \bibinfo {pages} {023408}
  (\bibinfo {year} {2014})}\BibitemShut {NoStop}%
\bibitem [{\citenamefont {Pfeiffer}\ \emph {et~al.}(2013)\citenamefont
  {Pfeiffer}, \citenamefont {Bell}, \citenamefont {Beck}, \citenamefont
  {Mashiko}, \citenamefont {Neumark},\ and\ \citenamefont
  {Leone}}]{Pfeiffer_Leone_2013PRA}%
  \BibitemOpen
  \bibfield  {author} {\bibinfo {author} {\bibfnamefont {A.~N.}\ \bibnamefont
  {Pfeiffer}}, \bibinfo {author} {\bibfnamefont {M.~J.}\ \bibnamefont {Bell}},
  \bibinfo {author} {\bibfnamefont {A.~R.}\ \bibnamefont {Beck}}, \bibinfo
  {author} {\bibfnamefont {H.}~\bibnamefont {Mashiko}}, \bibinfo {author}
  {\bibfnamefont {D.~M.}\ \bibnamefont {Neumark}}, \ and\ \bibinfo {author}
  {\bibfnamefont {S.~R.}\ \bibnamefont {Leone}},\ }\href {\doibase
  10.1103/PhysRevA.88.051402} {\bibfield  {journal} {\bibinfo  {journal} {Phys.
  Rev. A}\ }\textbf {\bibinfo {volume} {88}},\ \bibinfo {pages} {051402}
  (\bibinfo {year} {2013})}\BibitemShut {NoStop}%
\bibitem [{\citenamefont {Li}\ \emph {et~al.}(2015{\natexlab{b}})\citenamefont
  {Li}, \citenamefont {Bernhardt}, \citenamefont {R.}, \citenamefont {Warrick},
  \citenamefont {Pfeiffer}, \citenamefont {Bell}, \citenamefont {Haxton},
  \citenamefont {McCurdy}, \citenamefont {Neumark},\ and\ \citenamefont
  {Leone}}]{Li_Leone_JPhysB2015}%
  \BibitemOpen
  \bibfield  {author} {\bibinfo {author} {\bibfnamefont {X.}~\bibnamefont
  {Li}}, \bibinfo {author} {\bibfnamefont {B.}~\bibnamefont {Bernhardt}},
  \bibinfo {author} {\bibfnamefont {B.~A.}\ \bibnamefont {R.}}, \bibinfo
  {author} {\bibfnamefont {E.~R.}\ \bibnamefont {Warrick}}, \bibinfo {author}
  {\bibfnamefont {A.~N.}\ \bibnamefont {Pfeiffer}}, \bibinfo {author}
  {\bibfnamefont {M.~J.}\ \bibnamefont {Bell}}, \bibinfo {author}
  {\bibfnamefont {D.~J.}\ \bibnamefont {Haxton}}, \bibinfo {author}
  {\bibfnamefont {C.~W.}\ \bibnamefont {McCurdy}}, \bibinfo {author}
  {\bibfnamefont {D.~M.}\ \bibnamefont {Neumark}}, \ and\ \bibinfo {author}
  {\bibfnamefont {S.~R.}\ \bibnamefont {Leone}},\ }\href@noop {} {\bibfield
  {journal} {\bibinfo  {journal} {J. Phys. B: At. Mol. Opt. Phys.}\ }\textbf
  {\bibinfo {volume} {87}},\ \bibinfo {pages} {013415} (\bibinfo {year}
  {2015}{\natexlab{b}})}\BibitemShut {NoStop}%
\bibitem [{\citenamefont {Chen}\ \emph {et~al.}(2013)\citenamefont {Chen},
  \citenamefont {Wu}, \citenamefont {Gaarde},\ and\ \citenamefont
  {Schafer}}]{Mette.Ken.2013.ATA.He.LIP}%
  \BibitemOpen
  \bibfield  {author} {\bibinfo {author} {\bibfnamefont {S.}~\bibnamefont
  {Chen}}, \bibinfo {author} {\bibfnamefont {M.}~\bibnamefont {Wu}}, \bibinfo
  {author} {\bibfnamefont {M.~B.}\ \bibnamefont {Gaarde}}, \ and\ \bibinfo
  {author} {\bibfnamefont {K.~J.}\ \bibnamefont {Schafer}},\ }\href {\doibase
  10.1103/PhysRevA.88.033409} {\bibfield  {journal} {\bibinfo  {journal} {Phys.
  Rev. A}\ }\textbf {\bibinfo {volume} {88}},\ \bibinfo {pages} {033409}
  (\bibinfo {year} {2013})}\BibitemShut {NoStop}%
\bibitem [{\citenamefont {Eckstein}\ \emph {et~al.}(2016)\citenamefont
  {Eckstein}, \citenamefont {Yang}, \citenamefont {Frassetto}, \citenamefont
  {Poletto}, \citenamefont {Sansone}, \citenamefont {Vrakking},\ and\
  \citenamefont {Kornilov}}]{EcksteinVrakking2016.N2}%
  \BibitemOpen
  \bibfield  {author} {\bibinfo {author} {\bibfnamefont {M.}~\bibnamefont
  {Eckstein}}, \bibinfo {author} {\bibfnamefont {C.-H.}\ \bibnamefont {Yang}},
  \bibinfo {author} {\bibfnamefont {F.}~\bibnamefont {Frassetto}}, \bibinfo
  {author} {\bibfnamefont {L.}~\bibnamefont {Poletto}}, \bibinfo {author}
  {\bibfnamefont {G.}~\bibnamefont {Sansone}}, \bibinfo {author} {\bibfnamefont
  {M.~J.}\ \bibnamefont {Vrakking}}, \ and\ \bibinfo {author} {\bibfnamefont
  {O.}~\bibnamefont {Kornilov}},\ }\href@noop {} {\bibfield  {journal}
  {\bibinfo  {journal} {Physical review letters}\ }\textbf {\bibinfo {volume}
  {116}},\ \bibinfo {pages} {163003} (\bibinfo {year} {2016})}\BibitemShut
  {NoStop}%
\bibitem [{\citenamefont {Haxton}\ \emph {et~al.}(2011)\citenamefont {Haxton},
  \citenamefont {Lawler},\ and\ \citenamefont {McCurdy}}]{HLM2011}%
  \BibitemOpen
  \bibfield  {author} {\bibinfo {author} {\bibfnamefont {D.~J.}\ \bibnamefont
  {Haxton}}, \bibinfo {author} {\bibfnamefont {K.~V.}\ \bibnamefont {Lawler}},
  \ and\ \bibinfo {author} {\bibfnamefont {C.~W.}\ \bibnamefont {McCurdy}},\
  }\href {\doibase 10.1103/PhysRevA.83.063416} {\bibfield  {journal} {\bibinfo
  {journal} {Phys. Rev. A}\ }\textbf {\bibinfo {volume} {83}},\ \bibinfo
  {pages} {063416} (\bibinfo {year} {2011})}\BibitemShut {NoStop}%
\end{thebibliography}%

%

\end{document}